\documentclass[12pt]{iopart}

\usepackage{xspace}
\usepackage{color}
\usepackage{graphicx}
\pdfoutput=1

\renewcommand{\vec}{\mathbf}

\newcommand{\braket}[2]{\xspace\ensuremath{\langle #1\mid #2\rangle}\xspace}

\newcommand{\ket}[1]{\xspace\ensuremath{\mid #1\rangle}\xspace}
\newcommand{\matel}[3]{\xspace{\langle #1\mid #2\mid #3\rangle}\xspace}

\definecolor{dblue}{rgb}{0,0,0.6}
\definecolor{dred}{rgb}{0.9,0,0}
\definecolor{dgreen}{rgb}{0,0.4,0}

\begin{document}

\topical[ARPES measurements of the superconducting gap of Fe-based superconductors]{ARPES measurements of the superconducting gap of Fe-based superconductors and their implications to the pairing mechanism}

\author{P. Richard$^{1,2}$, T. Qian$^{1}$ and H. Ding$^{1,2}$}

\address{1 Beijing National Laboratory for Condensed Matter Physics, and Institute of Physics, Chinese Academy of Sciences, Beijing 100190, China}
\address{2 Collaborative Innovation Center of Quantum Matter, Beijing, China}
\eads{\mailto{p.richard@iphy.ac.cn}}

\begin{abstract}
Its direct momentum sensitivity confers to angle-resolved photoemission spectroscopy (ARPES) a unique perspective in investigating the superconducting gap of multi-band systems. In this review we discuss ARPES studies on the superconducting gap of the high-temperature Fe-based superconductors. We show that while Fermi-surface-driven pairing mechanisms fail to provide a universal scheme for the Fe-based superconductors, theoretical approaches based on short-range interactions lead to a more robust and universal description of superconductivity in these materials. Our findings are also discussed in the broader context of unconventional superconductivity.
\end{abstract}



\maketitle

\section{Introduction}

The year 2008 discovery of superconductivity in a Fe-based material with a critical temperature $T_c$ of 26 K \cite{Kamihara_JACS2008} was received by the community as a powerful stimulant. Even though the cuprate superconductors remain until now the absolute champions of high-temperature superconductivity, they are no longer alone. As with the cuprates, the Fe-based superconductors have layered structures, relatively high $T_c$'s and a proximity to magnetic instabilities that quickly earned them the label ``high-temperature superconductors". With the hope that they will provide key insights into high-$T_c$ superconductivity, these materials have been investigated intensively over the last few years and their study is now one of the most active field in condensed matter physics \cite{JohnstonAdv_Phys2010,Stewart_RMP2011}. 

Arguably the most important interrogation raised by this new class of materials that are the Fe-based superconductors is: What is their superconducting (SC) pairing mechanism? Many approaches can be used to address this issue. However, the most direct one is to investigate how the electronic structure evolves as the system enters the SC state. Any superconductor gains energy upon entering the SC state by opening an energy gap at the Fermi surface (FS) of its electronic structure. In fact, this SC gap is the proper order parameter characterizing superconductivity. While conventional superconductors exhibit a uniform SC gap all over their FS, unconventional pairing mechanisms may lead to more exotic momentum dependence of the amplitude and phase of the SC gap. As an example, the cuprate superconductors are known as $d$-wave gap materials, with a SC gap that has nodes in the momentum space, which imposes severe restrictions to the theories used to describe the SC pairing mechanism. 

Angle-resolved photoemission spectroscopy (ARPES) is a momentum-resolved probe with sufficient energy resolution to determine precisely the SC gap of materials in the momentum space. It is thus a tool of choice for investigating the SC gap and the electronic structure of materials that have a multi-band FS like the Fe-based superconductors. A few reviews of ARPES results on Fe-based are available in the literature \cite{RichardRoPP2011,Y_Huang_AIP2012,Kordyuk_LTP38,ZH_LiuCPB22,ZR_YeCPB22}. The current one focusses on SC gap measurements of the Fe-based superconductors and closely related topics. Using data accumulated over the past 6 years, we show how ARPES is used to provide crucial information on the pairing mechanism.

In the next chapter, we introduce the general reader to the basic principles of ARPES and to its use in the study of Fe-based superconductivity. Then follows a chapter in which we define the SC gap and introduce the reader to the different theoretical approaches that can be used in trying to understand the SC pairing mechanism. Two popular approaches, namely the quasi-nesting model and the $J_1$-$J_2$-$J_3$ models, are exposed in Chapters \ref{section_QN_model} and \ref{section_strong_coupling}, respectively. In particular, we demonstrate how the FS topology cannot be the driving force for the pairing mechanism in this family of superconductors, which is more consistent with short-range interactions. Preceding the concluding remarks, we devote one chapter to the determination of the phase of the SC gap (Chapter \ref{section_phase}), one chapter on nodes and SC gap anisotropy (Chapter \ref{section_nodes}), as well as one chapter on the role of the orbital degree of freedom (Chapter \ref{section_orbital}).

\section{Introduction to ARPES and electronic structure of Fe-based superconductors}

\subsection{Basic principles of ARPES}

\begin{figure}[!t]
\begin{center}
\includegraphics[width=14 cm]{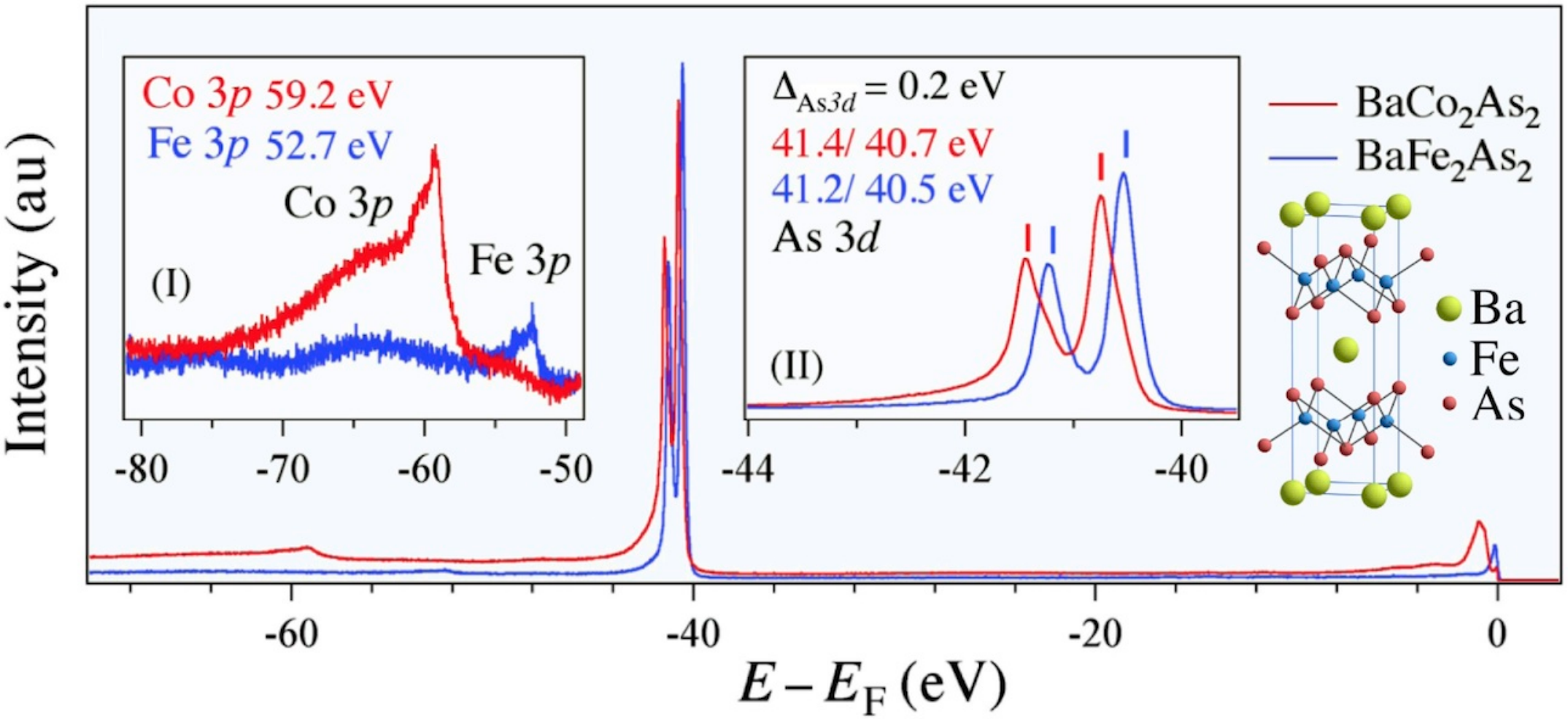}
\end{center}
\caption{\label{Fig_core_Co}(Colour online). Core level spectra of BaCo$_2$As$_2$ (red) and BaFe$_2$As$_2$ (blue) recorded with 195 eV photons. Insets I and II are zooms on the Fe/Co 3$p$ and  As 4$d$ levels, respectively. Reprinted with permission from \cite{Nan_XuPRX3}, copyright \copyright\xspace (2013) by the American Physical Society.}
\end{figure}

ARPES is an advanced photoemission spectroscopy (PES) technique. In PES [which includes ultraviolet PES (UPS) and x-ray PES (XPS)], we measure the kinetic energy of electrons emitted from the surface of a sample under the excitation of a photon flux of known energy $h\nu$ and vector potential $\vec{A}$. According to the conservation of energy and assuming that the photoemission process can be decomposed into three independent steps, namely (i) \emph{the excitation of  the initial state $\ket{i}$ into a bulk final state with no interaction between the excited electron and the core hole created (sudden approximation)}, (ii) \emph{the electron travel into the material} and (iii) \emph{the escape into a final state $\ket{f}$ through the surface potential}, the kinetic energy $E_k$ of these photoemitted electrons is the same as the energy $\varepsilon(k)$ relative to the Fermi level ($E_F$) they had before the photoemission process, \emph{modulo} a constant called the work function $\phi$, which represents the energy necessary to overcome the surface potential. PES is used as a fingerprint of the elemental constitution and chemical environment of the materials probed. As an example, we compare in Fig. \ref{Fig_core_Co} the PES shallow core level spectra of BaFe$_2$As$_2$ and BaCo$_2$As$_2$, which share the same crystal structure. The spectra show peaks characteristic of the Fe $3p$ and Co $3p$ states at different energy positions, indicating the different elemental compositions of these two compounds. Even though both materials contain As at the same crystal sites and in the same proportion, the different chemical environments and the different carrier concentrations resulting from the different electronic 3$d$ band fillings of Fe$^{2+}$ and Co$^{2+}$ lead to a shift in the energy positions of the As $3d$ core levels.

\begin{figure}[!t]
\begin{center}
\includegraphics[width=14 cm]{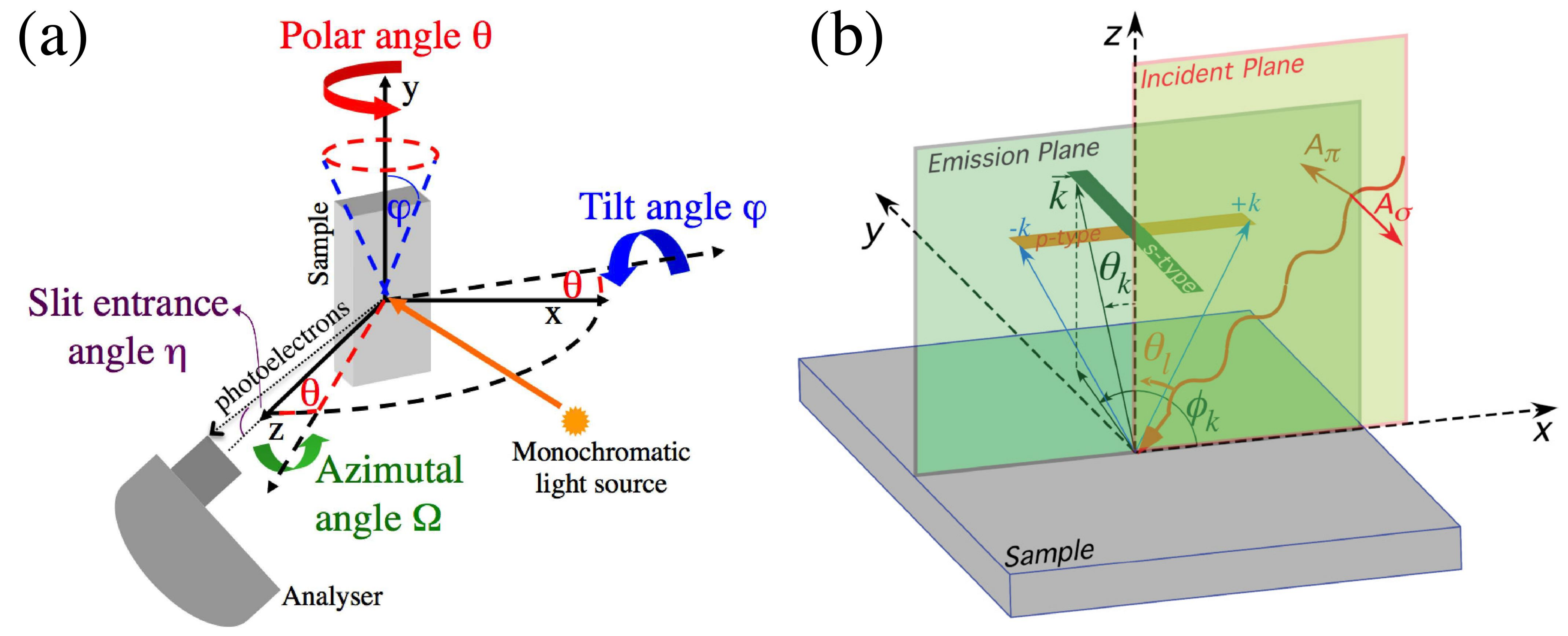}
\end{center}
\caption{\label{Sigma_pi}(Colour online). (a) Definitions of the angles used in the ARPES measurements. (b) Definitions of the $\pi$ and $\sigma$ configurations, along with the various angles used in the calculations. (b) Reprinted with permission from \cite{XP_WangPRB85}, copyright \copyright\xspace (2012) by the American Physical Society.}
\end{figure}

Because they usually disperse, the electronic states near $E_F$ cannot be uniquely represented by their energies. In addition to the conservation of energy, ARPES takes advantage of the conservation of the in-plane momentum by measuring the direction of emission of the photoemitted electrons, which is controlled by the relative orientation of the sample surface and the detector. Figure \ref{Sigma_pi}a illustrates the configuration mostly used nowadays in ARPES measurements, in which the detector position is fixed but the sample orientation can be moved from the normal emission direction by a polar angle $\theta$ and a tilt angle $\varphi$. The momenta corresponding to the photoemitted electrons can thus be simply expressed as a function of $\theta$ and $\varphi$:

\begin{center}
\begin{equation}
k_x=\frac{\sqrt{2mE_{kin}}}{\hbar}\sin\theta,\hspace{1cm} k_y=\frac{\sqrt{2mE_{kin}}}{\hbar}\cos\theta\sin\varphi \label{eq_ARPES_simple}
\end{equation}
\end{center}

In practice, modern semi-hemispherical energy analysers allow simultaneous measurements of the kinetic energy of electrons corresponding to different momenta inside a certain acceptance angle $2\eta_{max}$. The wider the acceptance angle, the larger the portion of the momentum space covered in a single measurement. We commonly call the vertical slit configuration and the horizontal configuration the configurations for which the detector slit is parallel and perpendicular to the polar rotation vector, respectively. Naming $\eta$ the slit entrance angle corresponding to the relative angle between the normal to the analyser slit and the direction of the electrons entering through the slit, the angle-momentum transformations become:

\begin{eqnarray}
k_x&=\frac{\sqrt{2mE_{kin}}}{\hbar}\sin\theta\cos\eta, \nonumber\\
k_y&=\frac{\sqrt{2mE_{kin}}}{\hbar}(\cos\theta\sin\varphi\cos\eta+\cos\varphi\sin\eta) \label{eq_ARPES_full_V}
\end{eqnarray}

\noindent for the vertical slit configuration and:

\begin{eqnarray}
k_x&=\frac{\sqrt{2mE_{kin}}}{\hbar}(\sin\theta\cos\eta +\cos\theta\sin\eta), \nonumber\\
k_y&=\frac{\sqrt{2mE_{kin}}}{\hbar}(\cos\theta\sin\varphi\cos\eta-\sin\theta\sin\varphi\sin\eta) \label{eq_ARPES_full_H}
\end{eqnarray}

\noindent for the horizontal slit configuration. 

The ARPES signal $I(\vec{k},E,\vec{A},h\nu)$ is proportional to the one-particle spectral weight $A(\vec{k}, E)$, which is the probability to have an electron in the sample with momentum $\vec{k}$ and energy $E$, times the Fermi-Dirac distribution $f(E,T)$: 
	
\begin{equation}
\label{eq_ARPES}
I(\vec{k},E,\vec{A},h\nu) = |M(\vec{k},E,\vec{A},h\nu)|^2A(\vec{k},E)f(E,T)
\end{equation}

\noindent where  

\begin{equation}
M(\vec{k},E,\vec{A},h\nu)= M_{if}= \matel{f}{\vec{A}\cdot\vec{r}}{i}
\end{equation}

\noindent represents the photoemission matrix element determined by the photoemission process itself expressed in terms of the potential vector $\vec{A}$ and the position operator $\vec{r}$. Although $M$ carries no direct information on the band dispersion, it contains precious information on the nature of the electronic states probed. For example, the photoemission intensity of the near-$E_F$ states in Ba$_{0.6}$K$_{0.4}$Fe$_2$As$_2$ exhibits an anti-resonance profile at 56 eV corresponding to the Fe $3p$ absorption edge, thus indicating that these states mainly originate from Fe \cite{Ding_JPCM2011}. In addition to the elemental character, $M$ can also provide important information on the orbital nature of the electronic states studied if one considers very simple selection rules. Since $\mid M_{if}\mid^2$ is a scalar observable, it is possibly non-zero only if it transforms under crystal symmetry operations like the fully symmetric irreducible representation $\Gamma_1$ of the corresponding crystallographic group. This means that the decomposition of the tensor product of $\Gamma_i$, $\Gamma_f$ and $\Gamma_{op}$, which are the representations associated to \ket{i}, \ket{f} and $\vec{A}\cdot\vec{r}$, respectively, must contain $\Gamma_1$, which is possible only if their total parity is even. The plane wave $\braket{r}{f}=e^{i\vec{k}\cdot\vec{r}}$ is always an even state with respect to the emission plane, as defined in Fig. \ref{Sigma_pi}b. With respect to that same plane, the operator $\vec{A}\cdot\vec{r}$ has an even and a odd parity, respectively, for the so-called $\sigma$ and $\pi$ experimental configurations also illustrated in Fig. \ref{Sigma_pi}b. Using the proper set of coordinates, one can thus deduce the symmetry of the initial state from the knowledge of the parity of both $\vec{A}\cdot\vec{r}$ and the final state. 

\begin{figure}[!t]
\begin{center}
\includegraphics[width=14 cm]{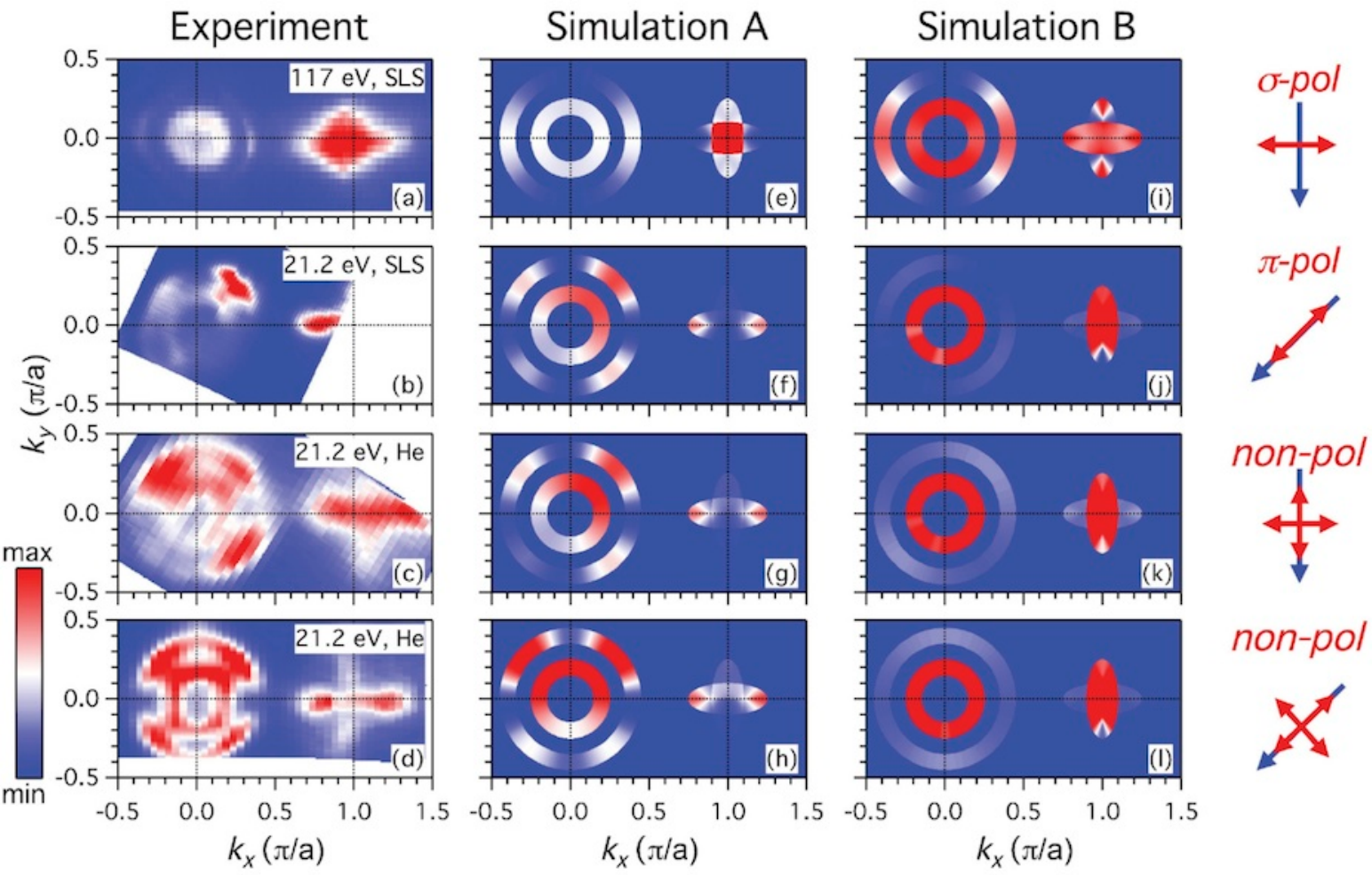}
\end{center}
\caption{\label{orbital_patterns}(Colour online). FS intensity patterns of Ba$_{0.6}$K$_{0.4}$Fe$_2$As$_2$. (a)-(d) Experimental results with different photon energies, polarizations and incident beam directions. (e)-(h) Corresponding simulated results using the simplified model described in Ref. \cite{XP_WangPRB85} (Simulation A: optimized orbital configuration). The inner $\Gamma$-centred $\alpha$ and $\alpha'$ FS pockets with $d_e$ and $d_o$ orbital characters are considered degenerate. The outer one ($\beta$ band) is associated to the $d_{xy}$ orbital. The tip of the M-centred FS pockets has pure $d_{xz}$ or $d_{yz}$ orbital characters while the inner part carries a dominant $d_{xy}$ orbital character. (i)-(l) Same as (e)-(h) but using a wrong orbital assignment (Simulation B). The orbital characters of the $\beta$ and $\alpha'$ bands have been exchanged compared to Simulation A. The orbital characters of the tip and inner part of the M-centred FS have also been exchanged. Red double-arrows and blue arrows indicate the in-plane components of the orientation of the light polarization and direction, respectively. Reprinted with permission from \cite{XP_WangPRB85}, copyright \copyright\xspace (2012) by the American Physical Society.}
\end{figure}

Exact calculations of the photoemission matrix elements are complicated and it is not always possible to go beyond simple selections rules. However, some attempts have been made to extract further information from the intensity patterns of the FS mappings. Using a simplified approach capturing the main angular dependence of the 3$d$ electronic orbitals, Wang \emph{et al.} \cite{XP_WangPRB85} established the main orbital distribution along the various FSs of the Fe-based superconductors. Fig. \ref{orbital_patterns} illustrates the comparison between the experimental FS patterns and simulated patterns for Ba$_{0.6}$K$_{0.4}$Fe$_2$As$_2$ under different experimental configurations. The experimental data show strongly anisotropic intensity patterns which are qualitatively well reproduced by Simulation A, which assumes a particular orbital distribution. In contrast, the agreement is rather bad for Simulation B, which assumes a different orbital distribution for the Fe 3$d$ states. As long as the matrix elements allow their observation, it is important to stress that the electronic dispersions measured experimentally are unaffected by the experimental setup. 

\subsection{Main advantages and limitations of the ARPES technique}

Usually, the near-$E_F$ electronic states in crystalline materials disperse in the momentum space, and thus necessitate a momentum-resolved characterization. ARPES is one of the only experimental probes available for this purpose. Moreover, the extraction of the information recorded by ARPES is arguably much easier to analyse than for other techniques. Unlike resonant inelastic x-ray scattering (RIXS \cite{Ament_RMP83}) for example, ARPES measures directly the single-particle spectral weight rather than transitions between two electronic states. In contrast to de Hass-van Alphen measurements \cite{Carrington_RoPP74}, which is also largely viewed as a powerful tool to measure the FS, ARPES does not require fit to theoretical models \emph{a priori}, and the raw data can be interpreted directly. This direct visualization of the momentum-resolved electronic states is a significant advantage when investigating multi-band materials. ARPES data are also obtained in the absence of external magnetic field perturbation and can be recorded even for relatively ``dirty" materials, for which the short electronic mean free path limits or even prevents the use of de Hass-van Alphen measurements. Actually, this situation often occurs in the study of high-temperature superconductors such as the cuprates and the Fe-based superconductors, for which doping is introduced through chemical substitution, thus inducing intrinsic disorder. Finally, it is worth emphasizing that ARPES is much more than a tool to access the FS of materials. Indeed, it can be used to determine the electronic structure over a wide energy range. This allows the measurement of momentum-resolved gaps, as well as the estimation of band renormalization related to electronic correlations.   

Despite its numerous advantages, ARPES, like any other experimental probe, also has its own limitations and comparison with other experimental techniques is sometimes either necessary or strongly encouraged. Although it can be viewed as an advantage when investigating surface phenomena such as in the study of the topological insulators, the surface sensitivity of ARPES is more often regarded as a disadvantage. In part for this reason, samples must be cleaved and measured in ultra-high vacuum conditions better than $10^{-9}$ Torr, which requires a complicated set of pumping stages. The better the vacuum, the longer the lifetime of the samples. Consequently, vacuum in the $10^{-11}$ Torr range are preferable and efforts are still devoted to the improvement of the vacuum conditions. The surface sensitivity of ARPES is also an obstacle when trying to access the bulk properties of materials. Nevertheless, the electronic states at the surface are always related to the bulk electronic states, a relationship qualitatively described by the equation $surface=bulk+\delta$. The reliability of the ARPES data as a measure of the bulk properties is thus directly related to the size of $\delta$, which varies from one compound to another. In practice, precious information can be deduced even when $\delta$ is large. Indeed, the surface states observed are often limited to a single chemical potential shift due to the polarity of the surface, which leaves the electronic structure almost intact, or to band foldings that are easy to identify. Several conditions help us to conclude that $\delta$ is small:

\begin{enumerate}
\item Low energy electron diffraction (LEED) pictures do not show obvious surface reconstruction;
\item The core levels of the relevant elements are not doubled;
\item The surface carrier doping, as determined from the Luttinger theorem, is consistent with that of the bulk;
\item The band dispersions are similar, albeit for some renormalization, to local density approximation (LDA) predictions;
\item The FS evolves smoothly with doping;
\item The electronic structure (band dispersion, gap size, etc...) varies with $k_z$, in sharp contrast to pure surface states;
\item The SC gap observed by ARPES closes at the bulk $T_c$;
\item No unexpected band folding is observed by ARPES;
\item Gaps measured by ARPES are consistent with gaps measured from bulk-sensitive probes. It is important to note that ARPES bulk-sensitivity is highly enhanced by the use of very low photon energies ($h\nu<9$ eV) \cite{KissPRL2005,SoumaRSI2007} or high photon energies ($h\nu>500$ eV) \cite{KamakuraEPL2004}.
\end{enumerate}

\begin{figure}[!t]
\begin{center}
\includegraphics[width=14 cm]{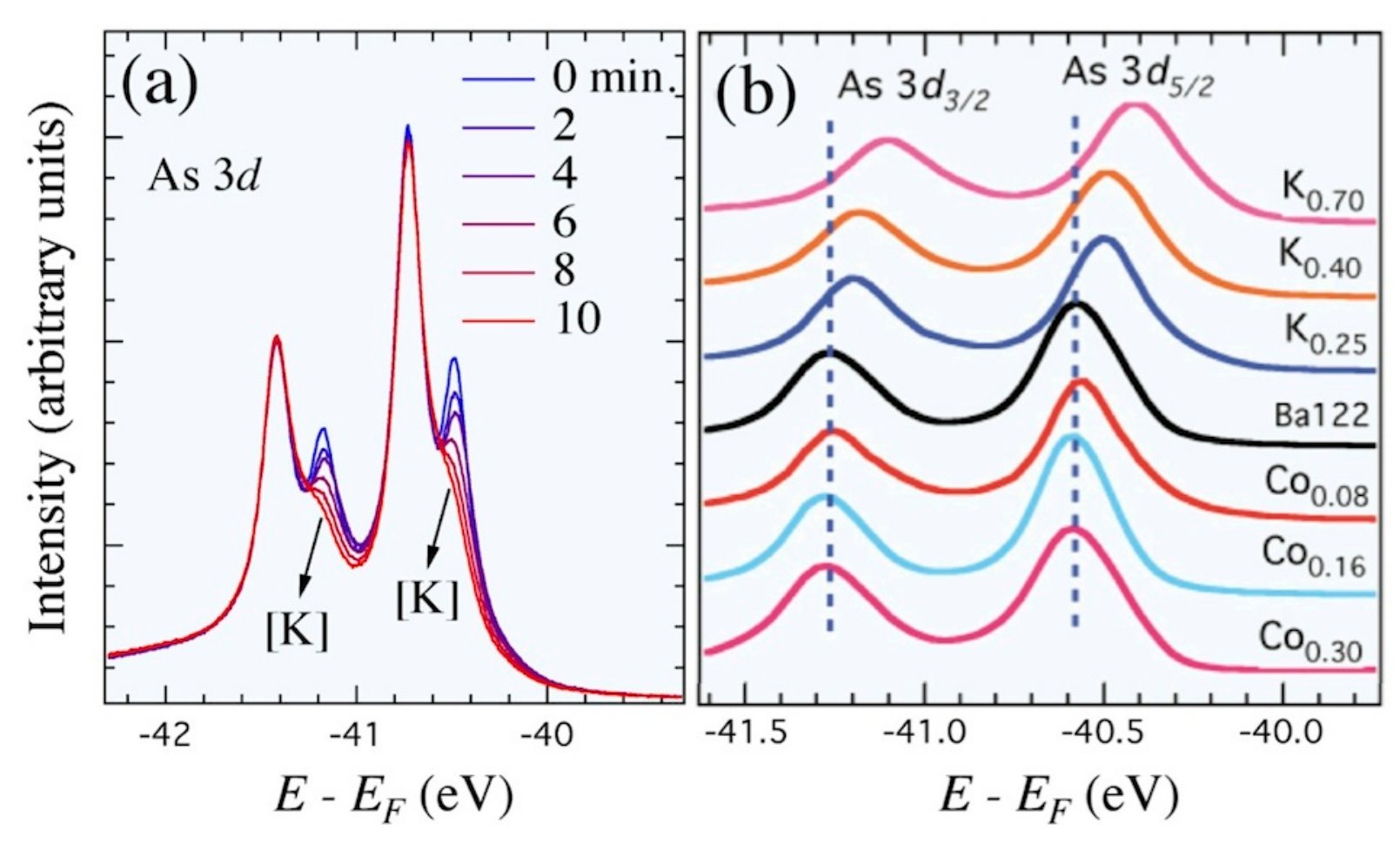}
\end{center}
\caption{\label{surface_core_As}(Colour online). (a) Evolution of the photoemission spectra of the As $3d$ core levels in EuFe$_2$As$_2$ as a function of the time of potassium evaporation. (b) As $3d$ core levels of the Ba$_{1-x}$K$_x$Fe$_2$As$_2$ and BaFe$_{2-x}$Co$_x$As$_2$ series. (a) Reprinted with permission from \cite{RichardJPCM26}, copyright \copyright\xspace (2014) by IOP Publishing. (b) Reprinted with permission from \cite{Neupane_PRB2011}, copyright \copyright\xspace (2011) by the American Physical Society.}
\end{figure}

Among all Fe-based superconductors, the 11-chalcogenide and 111-pnictide systems are in principle the most suitable to ARPES measurements because they lead to non-polar cleaved surfaces. Despite a band structure similar to that of other Fe-based superconductors, the 1111 system, on the other hand, leads to a strongly charged surface with a total FS volume incompatible with the sample composition \cite{DH_LuNature2008, Kondo_PRL2008, C_LiuPRB2010, HY_LiuPRL2010}. Particular attention must be devoted to the 122 system since it is by far the structure (illustrated in Fig. \ref{Fig_core_Co}) the most studied by ARPES. The cleavage of the sample occurs at the Ba plane. For electrostatic stability, half of the Ba remains on the cleaved surface, which is therefore a surface termination that differs from the bulk. Does that affect the electronic structure of the Fe-As layers situated below? Fortunately, an early LEED and STM study concluded in the absence of surface reconstruction in BaFe$_2$As$_2$ \cite{Nascimento_PRL2009}. However, a band folding leading to the emergence of photoemission intensity at the X $(\pi/2,\pi/2)$ point has been reported in SrFe$_2$As$_2$ \cite{Hsieh}, EuFe$_2$As$_2$ \cite{RichardJPCM26,Thirupathaiah_PRB84}, Ca$_{0.83}$La$_{0.17}$Fe$_2$As$_2$ \cite{YB_Huang_CPL30} and Ba(Fe$_{1-x}$Ru$_x$)$_2$As$_2$ \cite{Dhaka_prl2011}. The effect is relatively minor though and does not modify the main band dispersion. More serious is the recent report of a surface state affecting the As $3d$ and P $2p$ core levels in EuFe$_2$(As$_{1-x}$P$_x$)$_2$ \cite{RichardJPCM26}. Indeed, As and P are directly bounded to the Fe atoms mainly responsible for the FS of the Fe-based superconductors. We show in  Fig. \ref{surface_core_As}a the core levels of EuFe$_2$(As$_{1-x}$P$_x$)$_2$ under K evaporation \cite{RichardJPCM26}. Before evaporation, four peaks can easily be distinguished. As the time of evaporation increases, one pair of peaks associated to a surface state is slowly suppressed while the other pair, representative of the bulk, remains nearly unaffected. As shown in Fig. \ref{surface_core_As}b, such strong surface effect is not observed in Ba$_{1-x}$K$_x$Fe$_2$As$_2$ and BaFe$_{2-x}$Co$_x$As$_2$ \cite{Neupane_PRB2011}, for which systematic measurements of the SC gap have been done.

Due to the discontinuity at the surface of the samples measured, the component $k_z$ of the momentum perpendicular to the surface is not a good quantum number. This is a handicap when studying systems with tri-dimensional (3D) electronic structures. Nevertheless, there are a few ways in which ARPES can provide information on $k_z$ \cite{DamascelliPScrypta2004}. For the study of the Fe-based superconductors, the main approximation used to access the $k_z$ electronic dispersion is the nearly-free electron approximation, which is the logical extension of the 3-step model described above. Within this approximation, the energy $E_f$ of the final bulk states is simply described by:

\begin{equation}   
E_f=\frac{\hbar^2\vec{k}^2}{2m}-|E_0|=\frac{\hbar^2(\vec{k_{||}}^2+\vec{k}_{z}^2)}{2m}-|E_0|
\end{equation}

\noindent where $m$ is the free electron mass, $\vec{k_{||}}$ represents the in-plane component of the momentum and $E_0$ represents the bottom of the free electron energy dispersion. In the 3-step model the measured kinetic energy $E_{kin}$ of the photoemitted electrons corresponds simply to $E_f-\phi$. Defining the inner potential $V_0=|E_0|+\phi$, the momentum $k_z$ can be written as a function of $E_{kin}$ and the in-plane momentum $\vec{k_{||}}$:

\begin{equation}   
|\vec{k}_z|=\frac{\sqrt{2m}}{\hbar}\left[E_{kin}+V_0-\frac{\hbar^2\vec{k_{||}}^2}{2m}\right]^{1/2}
\end{equation}

\noindent In this formula, $V_0$ is an experimental parameter that can be adjusted to fit the $k_z$ periodicity of the experimental results. Keeping in mind that $E_{kin}$ is related to the photon energy $h\nu$ through the relation $E_{kin}=h\nu-\phi$, the variation of the momentum electronic dispersion along $k_z$ can be obtained by ARPES by tuning the photon energy. In the study of the Fe-based superconductors, such procedure has been first applied to Ba(Fe$_{1-x}$Co$_x$)$_2$As$_2$ \cite{VilmercatiPRB2009}, but it was then applied successfully to other Fe-based systems as well. The ARPES studies indicate that some bands show a non-negligible $k_z$ modulation whereas other bands do not disperse perpendicularly to the Fe-As layers. Although the value of the inner potential $V_0$ is compound-dependent, typical values around 15 eV are usually obtained in these systems \cite{RichardRoPP2011}. 

\subsection{Notation}

\begin{figure}[!t]
\begin{center}
\includegraphics[width=14 cm]{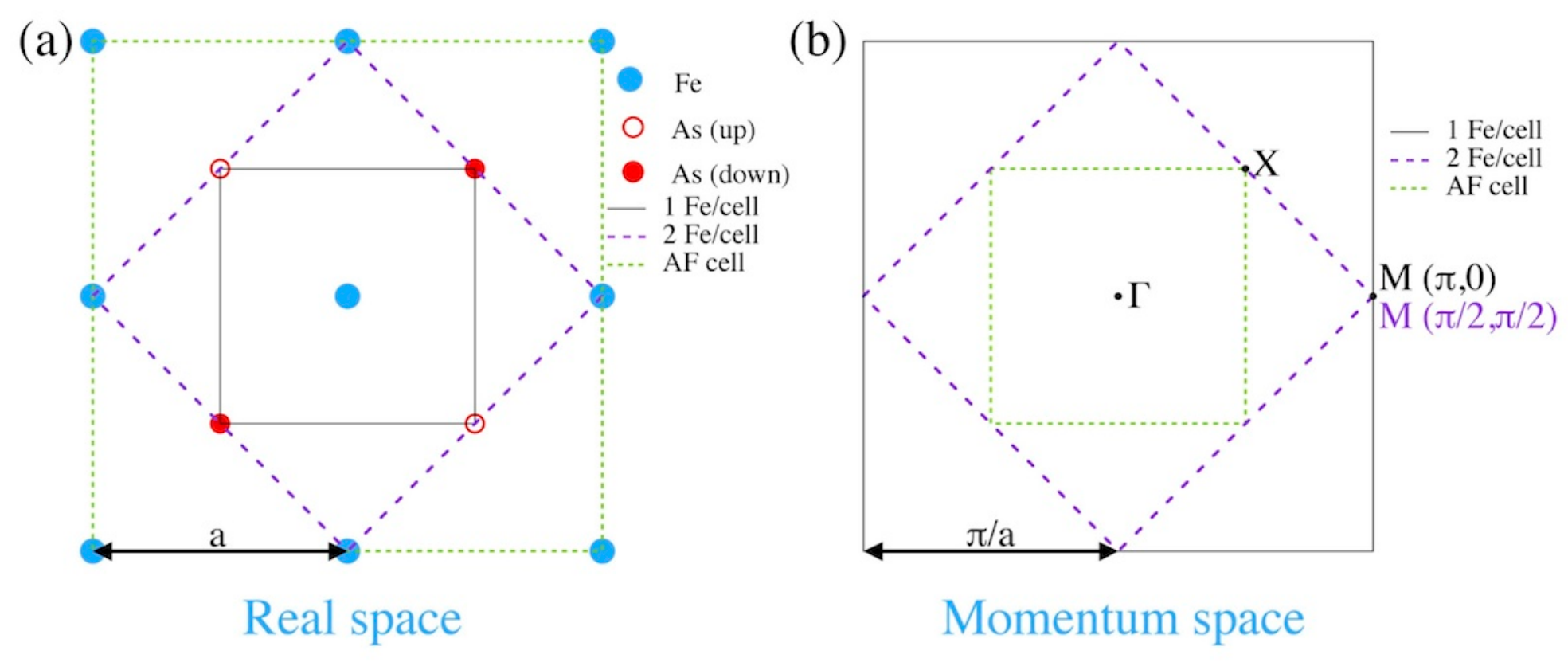}
\end{center}
\caption{\label{BZ_definition}(Colour online). (a) Top view of a Fe-As layer in the Fe-based superconductors, and definitions of the unit cells with 1 and 2 Fe atoms, as well as the AF unit cell. (b) Corresponding Brillouin zones in the momentum space, as well as the definitions of the symmetry points $\Gamma$, M and X.}
\end{figure}

The notation used in ARPES to describe the high-symmetry points of the first Brillouin zone (BZ) is not standard, and it is essential that we indicate the definitions that will be used in this review to describe the momentum space. We show in Fig. \ref{BZ_definition}a the top view of a typical Fe-As layer. Following a practice inherited from the study of the cuprate superconductors, where usually only the Cu atoms are represented, people often simplify their representation of the Fe-based materials by considering only the Fe atoms. Accordingly, one can define a unit cell containing a single Fe atom, with a lattice parameter $a$ coinciding with the distance between first Fe neighbours. The corresponding 1 Fe/unit cell BZ is illustrated in Fig. \ref{BZ_definition}b. In this notation, the zone centre is called $\Gamma$ and the zone boundary M $(\pi,0)$. Another point of interest, at $(\pi/2,\pi/2)$, is called X. 

However, because the As atoms do not lie in the Fe layer but are located alternatively in planes above and below the Fe layer, the real unit cell contains 2 Fe atoms, as illustrated in Fig. \ref{BZ_definition}a. As shown in Fig. \ref{BZ_definition}b, this leads to a 2 Fe/unit cell BZ that is half in size compared to the 1 Fe/unit cell BZ. In this alternative notation, the M point is now located at the $(\pi,\pi)$ corner of the BZ indexed in terms of the crystallographic lattice parameter $a'=\sqrt{2}a$. To add confusion, the notation for X and M is often swapped. Mainly for historical reasons, here we adopt the 1 Fe/unit cell description throughout this review paper, unless specified otherwise. In the presence of antiferromagnetic (AF) ordering, we can also define an AF unit cell and the corresponding AF BZ, as illustrated in Figs. \ref{BZ_definition}a and \ref{BZ_definition}b, respectively. Indeed, the AF ordering leads to band folding, as clearly evidenced in the parent compound of the 122 structural phase \cite{RichardPRL2010,LX_YangPRL2009,GD_LiuPRB2009,Kondo_PRB2010,M_YiPNAS2011,Y_KimPRB2011}.

Although the indexation of the BZ used to represent the ARPES results in terms of the 1 Fe/unit cell BZ or the 2 Fe/unit cell BZ does not affect the experimental results, it is nevertheless very important to keep in mind the implications of the real symmetry of the Fe-As layer. As we discuss later, the symmetry plays a crucial role in describing the electronic pairing. Moreover, the existence of As atoms alternating above and below the Fe planes may lead to confusion in the assignment of the orbital characters of the electronic states probed by ARPES \cite{CH_LinPRL107,Brouet_PRB86}.  

\subsection{Electronic structure of the Fe-based superconductors}
\label{section_electronic_structure}

All the Fe-based superconductors share the same basic structural blocs consisting in layers of Fe-Pn (Pn = P, As, Sb) or Fe-Ch (Ch = S, Se, Te) such as the ones of BaFe$_2$As$_2$ illustrated in Fig. \ref{Fig_core_Co}. Consequently, their electronic structures also share important similarities over a wide energy range, although details may vary from one compound to the other. As an archetype example, the electronic structure of Ba$_{0.6}$K$_{0.4}$Fe$_2$As$_2$ within 1 eV below $E_F$ is mainly composed of Fe 3$d$ orbitals, whereas the electronic states below, down to 5 or 6 eV, are mainly composed of As 4$p$ orbitals \cite{Ding_JPCM2011}. Except for a non-negligible band renormalization, which varies normally from 2 to 5 \cite{RichardRoPP2011}, LDA band structure calculations generally provide a good first approximation of the electronic band structure. In particular, early LDA calculations predicted that the Fe $3d$ bands should form 5 FSs \cite{Singh,F_Ma_PRB78,G_Xu_EPL2008}. Although this may depend on the precise electronic concentration, this is typically the case experimentally. Unlike the cuprates, the Fe-based superconductors are thus multi-band materials, and the characterization of their electronic structure is fundamentally non-trivial and requires experimental probes capable of momentum resolution.

As with normal metallic compounds, it is widely believed that the electronic structure near $E_F$ controls the electronic behaviour of the Fe-based superconductors. Actually, the FS topology of these materials is quite interesting: while hole-like pockets are generally observed around the $\Gamma$ point, electron-like pockets are normally found at the M point, which gives rise to the quasi-nesting model described in section \ref{section_QN_model}. The core of the problem of high-temperature superconductivity in the Fe-based superconductors consists in determining whether or not the FS topology plays a dominant role in the pairing mechanism. Using ARPES studies of the FS topology and of the SC gap, one of the main aims of the current topical review is to demonstrate how the FS topology of these systems cannot provide a universal picture for their SC pairing mechanism.

\section{The superconducting gap}
\label{section_gap_measurements}

\subsection{Definition of the superconducting gap}
\label{section_gap_definition}

The SC gap, defined by an amplitude and a phase, is the order parameter characterizing the SC state. Because it can access the electronic structure not only at the FS but also below, ARPES can measure the momentum-resolved SC gap. Strictly speaking though, ARPES can only access the amplitude of the SC gap directly, which is the main topic of this chapter. Nevertheless, such knowledge is very useful and can be used to test the validity of the theoretical models used to describe Fe-based superconductivity. In order to avoid possible confusion, here we define how SC gaps are evaluated from ARPES data.

In the framework of the BCS theory \cite{BCS_paper}, electron-hole mixing leads to the formation of two energy dispersions which are symmetrical with respect to $E_F$. In terms of the normal state dispersion $\epsilon(k)$ and the SC gap $\Delta(k)$, these Bogoliubov dispersions $E_{\pm}(k)$ describing the system below the critical temperature $T_c$ are characterized by the relation: 

\begin{equation}
\label{gap_BCS}
E_{\pm}(k)=\pm\sqrt{\epsilon^2(k)+\Delta^2(k)}
\end{equation}

\noindent and the corresponding spectral function $A(\vec{k},\omega)$ corresponds to 

\begin{equation}
\label{A_BCS}
A(\vec{k},\omega)=\frac{1}{\pi}\left\{\frac{|u_k|^2\Sigma''}{[\omega-E(k)]^2+\Sigma''^2}+\frac{|v_k|^2\Sigma''}{[\omega+E(k)]^2+\Sigma''^2} \right\},
\end{equation}

\noindent where $\omega$ is the energy relative to $E_F$, $\Sigma''$ is the linewidth broadening and $u_k$ and $v_k$ are the SC coherence factors defined as

\begin{equation}
|u_k|^2=1-|v_k|^2=\frac{1}{2}\left[1+\frac{\epsilon(k)}{E(k)}\right].
\end{equation}

\begin{figure}[!t]
\begin{center}
\includegraphics[width=12 cm]{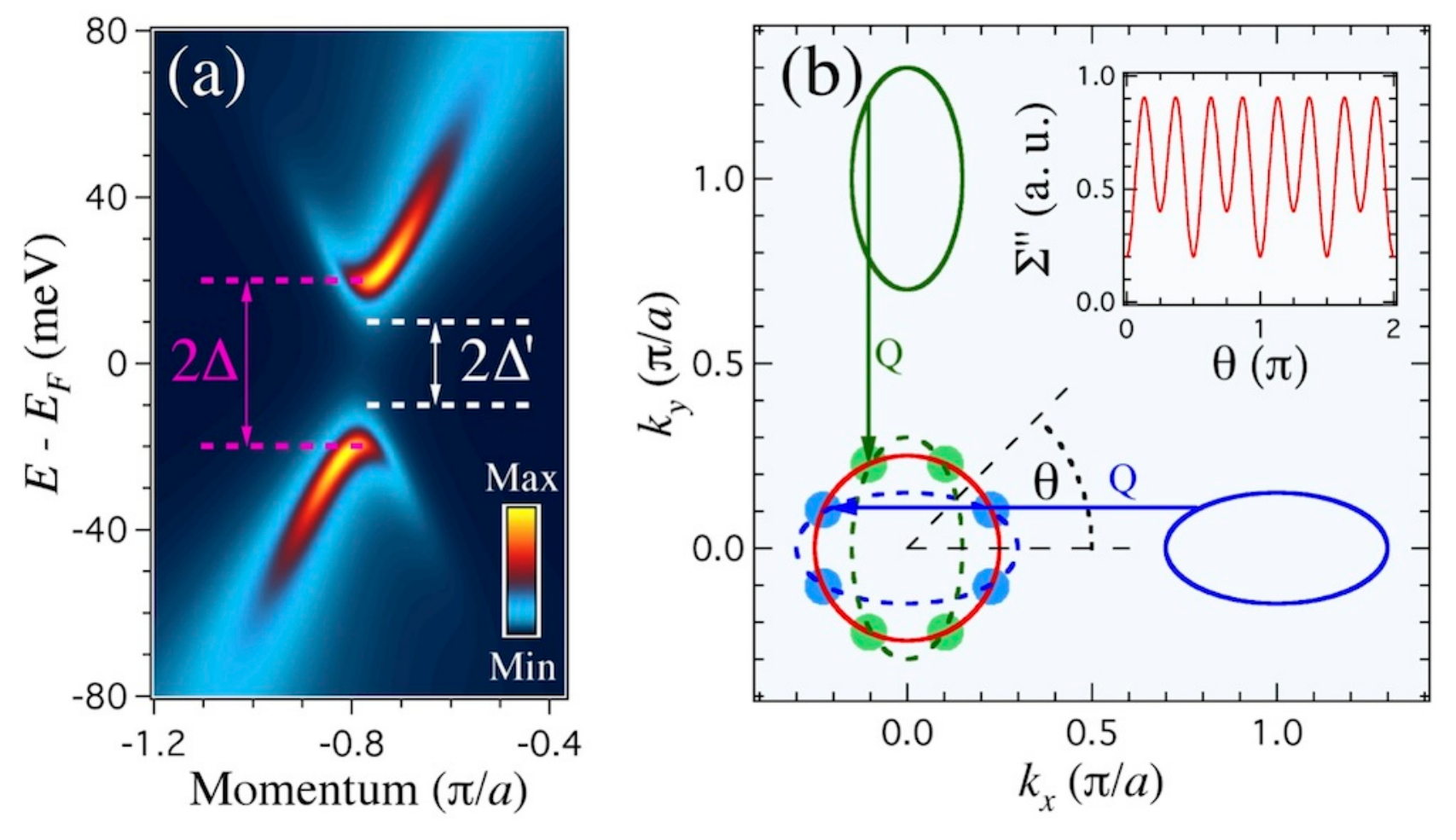}
\end{center}
\caption{\label{Gap_definition}(Colour online). (a) Simulation of the spectral function $A(\vec{k},\omega)$ in the presence of a 20 meV SC gap. We introduced an imaginary part to the self-energy with a quadratic dependence on energy in order to make the simulation more realistic. $\Delta$ corresponds to the SC gap while $\Delta^{\prime}$ is associated to an effective gap as would be measured by probes sensitive to a residual density-of-states. (b) Schematic FS of an hypothetical 2-band Fe-based superconductor. The dashed-line FSs have been translated by the AF wave vector $\vec{Q}$ to show where to expect stronger scattering (green and blue spots). The inset shows the schematic angular dependence of the imaginary part of the self-energy associated to interband scattering. Reprinted with permission from \cite{Y_Huang_AIP2012}, copyright \copyright\xspace (2012) by the American Institute of Physics.}
\end{figure}

Fig. \ref{Gap_definition}a simulates $A(\vec{k},\omega)$ in the SC state, with a $\Delta=20$ meV gap size. As indicated by a mark, this value corresponds to the maximum of the electronic dispersion, or equivalently, to the minimum gap location. Although ARPES cannot access the unoccupied states above $E_F$ for more than a few $k_BT$'s, it can easily track the band dispersion below $E_F$ and thus provide directly an accurate value for $\Delta$. Because this value is directly involved in Eq. \ref{gap_BCS}, we call $\Delta$ the SC pairing gap. To avoid any effect due to thermal broadening, the ARPES data are often ``symmetrized". This procedure exploits the electron-hole symmetry of the spectral function at the Fermi wave vector $\vec{k}_F$, \emph{i. e.} $A(\vec{k}_F,\omega)=A(\vec{k}_F,-\omega)$. As a consequence, the symmetric counterpart of $f(\omega,T)A(\vec{k},\omega)$ with respect to $E_F$ is simply $f(-\omega,T)A(\vec{k},-\omega)=[1-f(\omega,T)]A(\vec{k},\omega)$, which means that the effect of the Fermi function is removed from their sum. This method is very useful to visualize SC gaps. In practice though, unless the size of the gap is very small compared to the broadness of the quasiparticle peak, the values of $\Delta$ extracted from symmetrized and unprocessed data are almost the same. 

As illustrated in Fig. \ref{Gap_definition}a, the finite lifetime of the quasiparticles introduces a band broadness, which has several consequences on the interpretation of the SC gap. Whatever the origin of the scattering $\Sigma''$ leading to this broadening, the spectral function always shows a tail that extends inside the pairing gap. Therefore, alternatively to the pairing gap $\Delta$ defined as a gap in the electronic dispersion, one can define a gap $\Delta'$ corresponding to a gap in the density-of-states (DOS). As shown in Fig. \ref{Gap_definition}a, we necessarily have $\Delta'<\Delta$. An important corollary to this remark is that any experimental probe sensitive to the DOS would track $\Delta'$ rather than $\Delta$.  

The form of $\Sigma''(\vec{k})$ is not always trivial. Although an isotropic contribution is usually expected for impurity scattering, interband scattering is strongly dependent on the size and shape of the various FSs. This effect is illustrated in Fig. \ref{Gap_definition}b. In this example, 8 hot spots corresponding to $\vec{k}_F$ locations with stronger interband scattering are expected. Consequently, even in the presence of an isotropic pairing gap $\Delta$, it is possible to find an anisotropic DOS gap $\Delta'$ that reflects the ``undesired" influence of scattering.  

Besides the minimum gap location method describe above, other techniques are sometimes used to determine the SC gap of materials from ARPES data. One of them consists in evaluating the shift of the leading edge, called leading edge shift or leading edge gap (LEG). Obviously, the opening of a leading edge gap LEG below $T_c$ is a clear indication of a SC state. However, the LEG does not track the exact value of the pairing gap and it is necessarily smaller than $\Delta$. More importantly, it does not necessarily track the momentum dependence of $\Delta$ either. Indeed, the position of the LEG depends not only on $\Delta$, but also on $\Sigma''(\vec{k})$, and thus the momentum dependence of the LEG is more consistent with that of the DOS gap than that of the pairing gap \cite{Y_Huang_AIP2012}. Another major disadvantage of the LEG method in the study of a multi-band system is the spectral contamination from bands closely located in the momentum space. Actually, this latter aspect also affects the determination of the SC gap from fit to some spectral functions such as the Dynes function \cite{Dynes_PRL41}. In general, the use of such function in the estimation of the pairing gap is justified only in the presence of strong and sharp coherent SC peaks, the fits being mainly controlled by the position of the leading edge when these peaks are small or inexistent, thus modulating the momentum dependence of the estimated gap size. 

\subsection{Choosing a model for the superconducting pairing in the Fe-based superconductors}

Conventional superconductors are well described by the BCS theory \cite{BCS_paper}. In this theory, itinerant electrons are paired through electron-phonon interactions. Because the Cooper pairs are formed by electronic carriers with opposite spin and opposite momentum, it is somehow more convenient to describe the SC pairing mechanism in the momentum space. However, the electron-phonon interactions are not suitable to explain the electron pairing in unconventional superconductors such as the Fe-based superconductors. For these materials, it is widely believed that the interactions between electrons are sufficient to lead to the formation of Cooper pairs. How we derive the electronic structure and the electronic interactions should thus be related to whether the electronic pairing is naturally explained in the real space or in the momentum space. Although the space and momentum representations are simply related by a Fourier transform, and thus both representations are technically valid, the philosophical implications derived from each representation are very different. 

On one side, some calculation techniques use free electrons as starting point, and introduce a periodic potential representing the effect of the lattice on these electrons. Consequently, the electrons are ``weakly coupled" to the ions forming the lattice. The corresponding wave functions are usually called Bloch states. Such computation tools are best represented by the density function theory (DFT) methods, such as LDA. On the other side, the tight-binding method starts with local wave functions called Wannier functions, which are by definition ``strongly coupled" to the ions forming the lattice. The momentum dispersion is obtained from the overlap of the Wannier functions on neighbouring sites. By extension, we call ``weak coupling" theory a theory that describes naturally in the momentum space the properties of itinerant electrons, which are located in a narrow energy range near $E_F$, and we call ``strong coupling" theory a theory for which the relevant interactions are defined in the real space, over a few inter-atomic distances. Of course, in many practical cases, the physical systems are neither describe simply by a weak coupling theory or by a strong coupling theory. For a single-band system, this is well illustrated by the Hubbard Model:

\begin{equation}
\label{Hubbard_eq}
H=-\sum_{<i,j>}t_{i,j}(c_{i\sigma}^{\dagger}c_{j\sigma}+c_{j\sigma}^{\dagger}c_{i\sigma})+U\sum_in_{i\uparrow}n_{i\downarrow}
\end{equation}

\noindent where the first and second terms represent the kinetic energy and the single-site potential energy, respectively. In the first term of this equation, $t_{i,j}$ represents the energy for hopping between the sites $i$ and $j$, and $c_{i\sigma}^{\dagger}$ ($c_{i\sigma}$) is the creation (annihilation) operator for and electron of spin $\sigma$ at site $i$. The summation is performed over all the $<i,j>$ pairs, with $i\neq j$. In the second term of eq. \ref{Hubbard_eq}, U is the on-site repulsion energy and $n_{i\sigma}$ is the number operator for electrons of spin $\sigma$ at site $i$.

The weak coupling in the Hubbard model corresponds to situations for which the kinetic energy is much larger than the on-site energy, \emph{i. e.} $U<<t$. In contrast, the strong coupling corresponds to cases where $U>>t$. In reality, there is a large range of possibilities between these two limits, usually referred to as ``intermediate coupling", where both $U$ and $t$ play an important role in describing the physical properties of correlated electron systems. In fact, this is possibly the case for Fe-based superconductors. To add to the complexity of the problem, the Fe-based superconductors are multi-band systems, and there is no rule stating that the different bands should be correlated in the same way. In such circumstances, there might be physical phenomena that are better described by a weak coupling approach while others are better explained in terms of a strong coupling theory, and there should be possibly other situations where it is necessary to analyze the system studied in terms of the intermediate coupling. Our goal in this topical is strictly limited to the study of the SC pairing mechanism.  

\begin{figure}[!t]
\begin{center}
\includegraphics[width=11 cm]{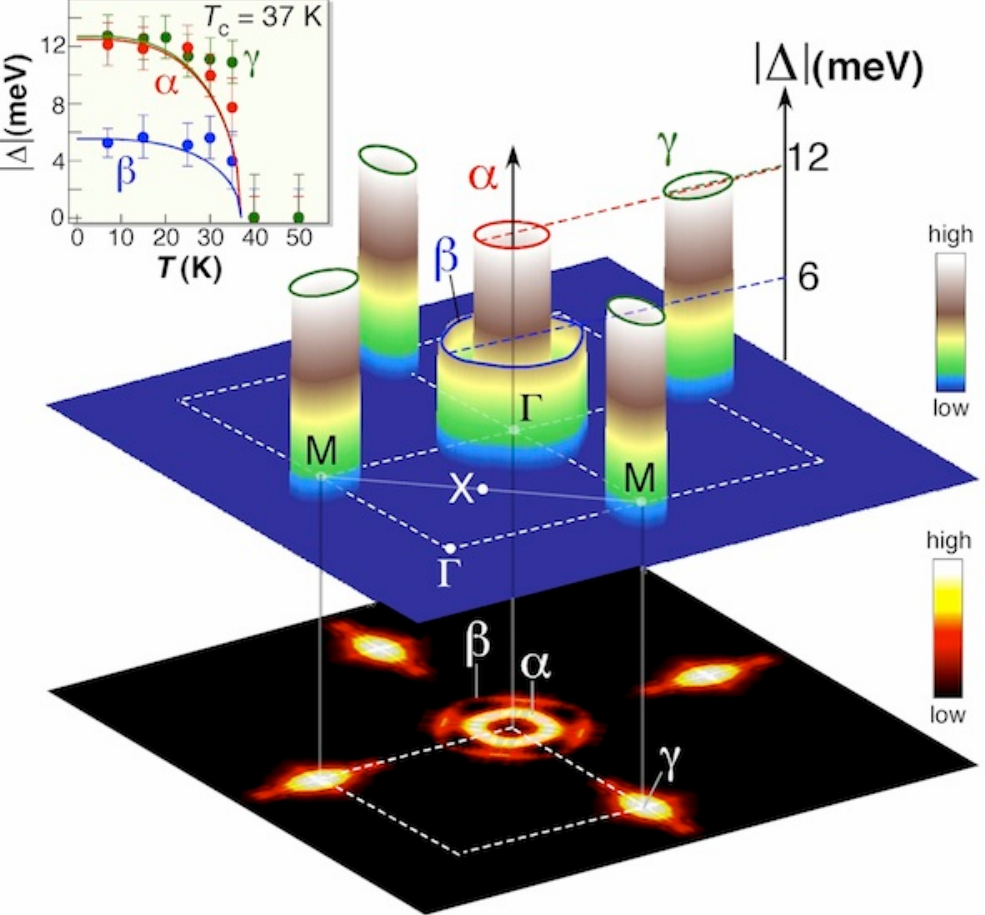}
\end{center}
\caption{\label{EPL_final}(Colour online). Three-dimensional plot of the SC gap size ($\Delta$) in Ba$_{0.6}$K$_{0.4}$Fe$_2$As$_2$ measured at 15 K on three FS sheets (shown at the bottom as an intensity plot) and their temperature evolutions (inset). Reprinted with permission from \cite{Ding_EPL}, copyright \copyright\xspace (2008) by the European Physical Society.}
\end{figure}

As bets and speculations were flourishing, the first ARPES reports on the SC gap in Ba$_{0.6}$K$_{0.4}$Fe$_2$As$_2$ \cite{Ding_EPL,L_Zhao}, summarized in Fig. \ref{EPL_final}, already established firmly the main characteristics of the SC gap of most Fe-based superconductors: this system shows a multi-gap structure, the gap amplitude is in the strong coupling regime and the SC gap on each FS sheet is either isotropic or weakly anisotropic. Further measurements confirmed these results and provided refinement of the gap structure around the M point \cite{Nakayama_EPL2009}.  

The next two chapters we compare to approaches to understand these results and the pairing: a weak coupling approach called the ``quasi-nesting model", and a strong coupling approach called the $J_1$-$J_2$-$J_3$ model. We will show that while the former one fails to provide a universal picture of the SC pairing mechanism, the latter one is so far as we can tell quite robust to the experimental observations. We caution that our conclusion on the nature of the pairing mechanism does not imply that all the physical phenomena in the Fe-based superconductors have to be described by strong coupling approach. It simply means that the pairing interactions occur over a distance that is equal or less than the distance between next-next Fe neighbours, and involve the electronic structure over an energy range significantly larger than the typical gap sizes measured.

\section{The quasi-nesting model}
\label{section_QN_model}

\subsection{Introduction to the quasi-nesting model}

The first ARPES observations were apparently consistent with the so-called quasi-nesting scenario \cite{Ding_EPL, MazinPhysicaC2009, Graser_NJP2009}, which is an extension of the notion of nesting. Pure nesting arises when large sections of the FS can be overlapped after a translation corresponding to a nesting vector $\vec{Q}_0$. In this circumstance the electronic system is unstable and usually develops a charge-density-wave (CDW) or a spin-density-wave (SDW) ordering characterized by $\vec{Q}_0$. In the case of quasi-nesting, these large portions of the FS do not overlap perfectly, but one can still define a vector $\vec{Q}$ at which the static susceptibility function $\chi_0(\vec{q}, E=0)$ exhibits a significant peak, indicating that the system is still prone to CDW or SDW ordering in the presence of weak interactions \cite{Cvetkovic_EPL2009}. In other words, two sections of FSs A and B are quasi-nested by the vector $\vec{Q}$ if for each $k_F$ positions of A we can find a $k_F$ location on section B such that $\vec{Q}+\delta\vec{q}_i$ connects the two points, with $|\delta\vec{q}_i|$ small. The robustness of the quasi-nesting conditions can be significantly reinforced when considering dynamical fluctuations and the dynamical susceptibility $\chi_0(\vec{q}, E)$ \cite{F_WangEPL2009}. In this case, not only the wave vector $\vec{Q}$ is allowed to fluctuate, but the energy as well, up to small variations $\delta E$. Obviously, such dynamical process is efficient only when hole-like FSs are quasi-nested with electron-like FSs.  

\begin{figure}[!t]
\begin{center}
\includegraphics[width=14 cm]{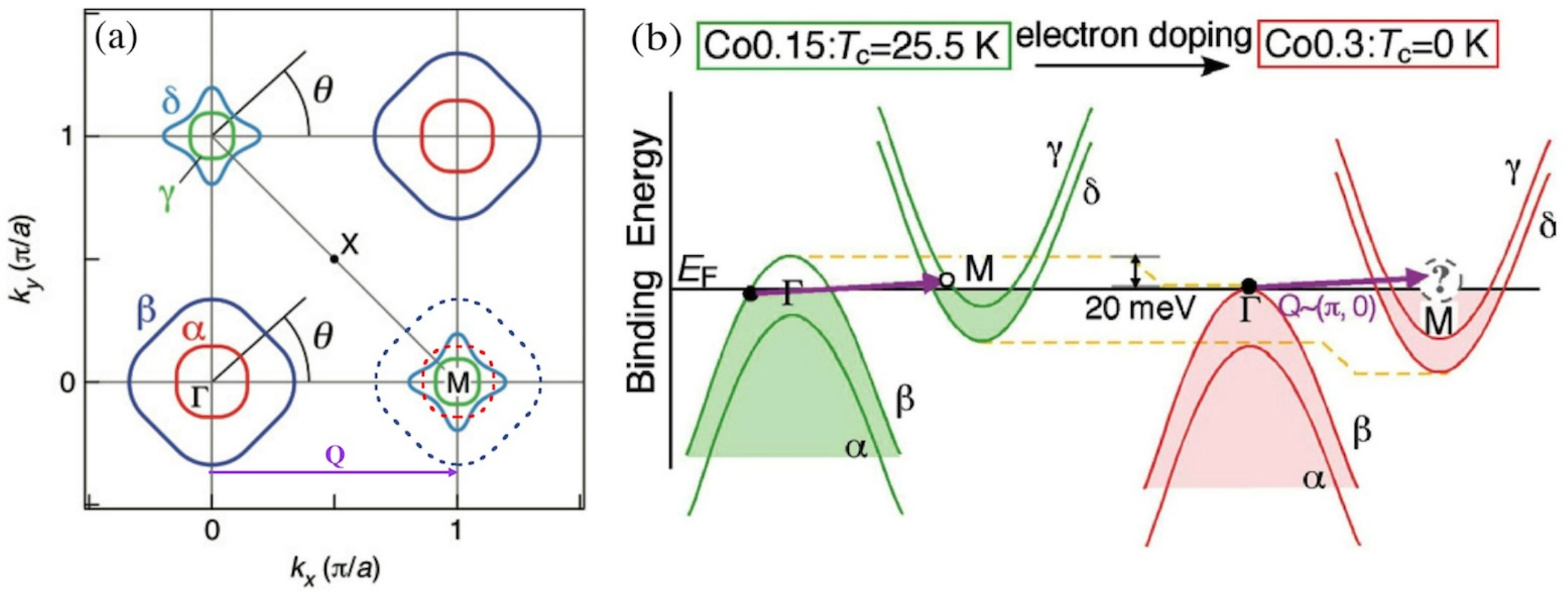}
\end{center}
\caption{\label{quasi_nesting}(Colour online). (a) FS of Ba$_{0.6}$K$_{0.4}$Fe$_2$As$_2$. The dashed lines correspond to FSs shifted by the AF wave vector $\vec{Q}$. (b) Illustration of the evolution of the quasi-nesting conditions in BaCo$_{2-x}$Fe$_{x}$As$_2$. Inter-band scattering is dramatically suppressed in the non-SC BaCo$_{1.7}$Fe$_{0.3}$As$_2$ sample since the hole-like $\alpha$ and $\beta$ bands at the $\Gamma$ point are basically occupied. Panel a is reprinted with permission from \cite{Nakayama_EPL2009}, copyright \copyright\xspace (2009) by the European Physical Society. Panel b is reproduced with permission from \cite{Sekiba_NJP2009}, copyright \copyright\xspace (2009) by IOP Publishing and Deutsche Physikalische Gesellschaft.}
\end{figure}

Using Ba$_{0.6}$K$_{0.4}$Fe$_2$As$_2$ as an example, we illustrate the notion of FS quasi-nesting in Fig. \ref{quasi_nesting}a. As detailed in the previous chapter, the FS of this material is composed by $\Gamma$-centred hole-like FSs and M-centred electron-like FSs. For comparison, we plot with dashed lines the hole-like FSs that have been shifted from $\Gamma$ to M by the AF vector $\vec{Q}$. In contrast to the size of the $\alpha$ FS, which is comparable to that of the electron-like FSs $\delta$ and $\gamma$, the size of the $\beta$ FS is much larger and therefore inter-band scattering involving the $\beta$ band is unlikely. Consistently, the gap amplitude determined experimentally was about 12 meV for all FSs except for the $\beta$ FS, on which a much smaller 6 meV SC gap was reported \cite{Ding_EPL,L_Zhao,Nakayama_EPL2009}, thus suggesting the importance of the FS topology. In support of this observation, anomalies in the electronic dispersion of bands that are quasi-nested were detected below $T_c$ \cite{RichardPRL2009}. Knowing the 12 meV energy size of the SC gap on the $\alpha$ FS and the electron-like FSs, the 25 meV energy of this anomaly is interpreted as an evidence for a 13 meV electron-mode coupling, which is in good agreement with the observation by inelastic neutron scattering of a 14 meV mode at the AF wave vector \cite{Christianson_Nature2008}.  

The best effective way to test the quasi-nesting scenario is to modulate the relative sizes of the $\Gamma$-centred hole-like FSs and M-centred electron-like FSs, which is done in practice by changing the electronic carrier concentration. The first attempt to check that with ARPES was done in a study of optimally-electron-doped BaCo$_{1.85}$Fe$_{0.15}$As$_2$ \cite{Terashima_PNAS2009}. Using the I$\alpha$ line of a He discharge lamp, Terashima \emph{et al.} showed that while the $\alpha$ FS does not cross $E_F$ at that particular photon energy, the $\beta$ FS shrinks to a size roughly matching the size of the expanding electron-like FSs at the M point, thus favouring inter-band scattering between the two sets of FSs. Interestingly, a strong coupling gap with $2\Delta_\beta/k_BT_c\approx 6$ was measured for the $\beta$ band, in sharp contrast with the weak coupling $2\Delta_\beta/k_BT_c\approx 3.7$ ratio measured in Ba$_{0.6}$K$_{0.4}$Fe$_2$As$_2$ \cite{Ding_EPL}. Unfortunately, due to in-plane doping leading to larger impurity scattering than for the off-plane doping of the Ba$_{1-x}$K$_{x}$Fe$_2$As$_2$ system, the coherence peak are ill-defined in the BaCo$_{2-x}$Fe$_{x}$As$_2$ series, and the systematic the evolution of the SC gap has never been studied by ARPES.

Further observations apparently consistent with the quasi-nesting model were also made for over-doped systems. Fig. \ref{quasi_nesting}b illustrates the particular situation in which the system is highly electron-doped and the tops of the $\Gamma$-centred hole-like FSs are band gapped. In this precise case, which corresponds to the FS topology of BaCo$_{1.7}$Fe$_{0.3}$As$_2$, electron-hole quasi-nesting is impossible \cite{Sekiba_NJP2009}. In apparent agreement with the quasi-nesting model, the $T_c$ of this compound vanishes. A similar observation has been reported for heavily hole-doped KFe$_2$As$_2$, in which the M-centred electron-like FSs are replaced by off-M-centred hole-like pockets, thus preventing electron-hole quasi-nesting \cite{Sato_PRL2009,Yoshida_JCPS72}. Accordingly, this material only has a small $T_c$ of 3 K. 

In fact, prior to the discovery of the 122-ferrochalcogenide superconductors, all Fe-based superconductors with a sufficiently high $T_c$ could be characterized by a FS formed by $\Gamma$-centred hole-like pockets and M-centred electron-like pockets, in support of the quasi-nesting model. In addition to the 122-ferropnictides, the 111-ferropnictides also satisfy this condition, and the magnitude of the large SC gaps reported indicates that the system is in the strong coupling regime. For instance a $2\Delta/k_BT_c\approx 8$ has been reported in NaFe$_{0.95}$Co$_{0.05}$As \cite{ZH_LiuPRB84}. While the nesting conditions are weakened in LiFeAs \cite{BorisenkoPRL2010} as compared to NaFe$_{0.95}$Co$_{0.05}$As \cite{ZH_LiuPRB84} and NaFeAs \cite{C_HePRL2010}, it is still fair to say that the hole-like and electron-like FSs pockets remain quasi-nested in the sense of the quasi-nesting concept described in this review. 

More challenging to the quasi-nesting approach was a series of theoretical calculations of non-quasi-nested FSs in Sr$_2$VFeAsO$_3$ \cite{KW_LeeEPL2010,MazinPRB2010_21311,Shein_JSNM2009,G_WangPRB2009}, which proved to be incompatible with the quasi-nested experimental FSs \cite{Qian_PRB2011}. Finally, the 11-chalcogenide FeTe$_{1-x}$Se$_{x}$ exhibits a similar FS topology \cite{Nakayama_PRL2010, Tamai_PRL2010,H_Miao_PRB2012} and large SC gaps as well \cite{Nakayama_PRL2010,H_Miao_PRB2012,Lubashevsky_NPhys2012}. Although the AF wave vector of the parent compound Fe$_{1+y}$Te does not coincide with the $\Gamma$-M wave vector, which leads to a folding of bands at the X point in their parent compound \cite{Y_XiaPRL2009}, appreciable neutron scattering at the $\Gamma$-M wave vector has been reported in SC samples of FeTe$_{1-x}$Se$_{x}$ \cite{BaoPRL102,S_LiPRB79,IikuboJPSJ2009,YM_QiuPRL2009,MookPRL104}. 

\subsection{Failure of the quasi-nesting model}
\label{section_weak_coupling}

\begin{figure}[!t]
\begin{center}
\includegraphics[width=12 cm]{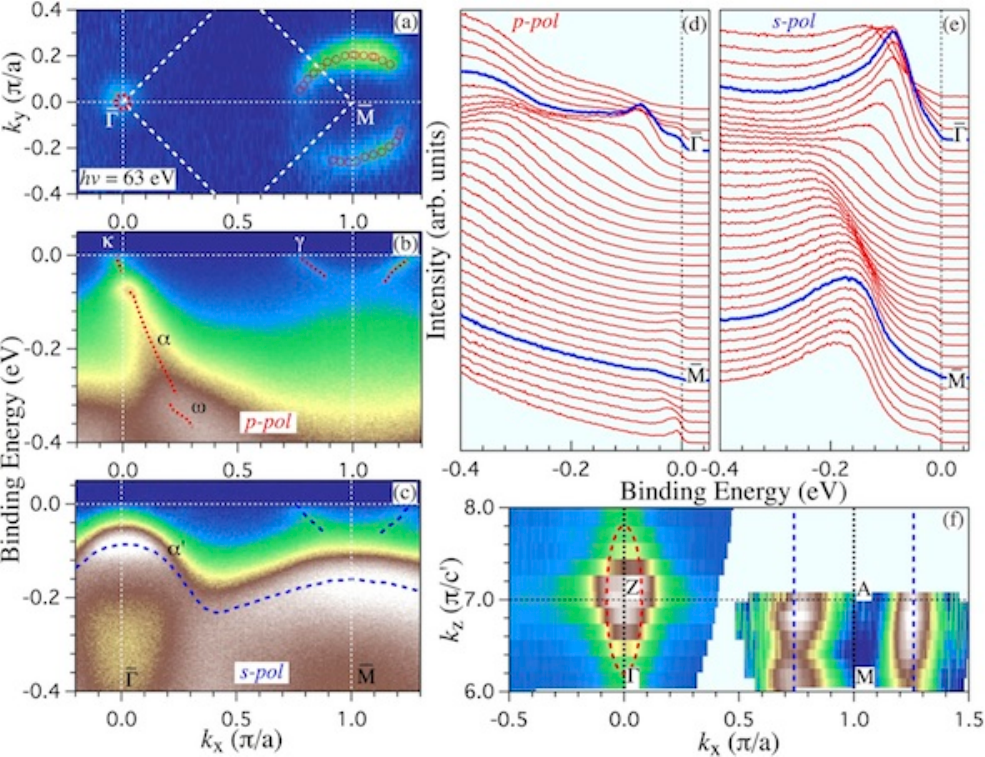}
\end{center}
\caption{\label{ch122FS}(Colour online). (a) ARPES FS intensity map of Tl$_{0.63}$K$_{0.37}$Fe$_{1.78}$Se$_2$ ($\pm 5$ meV integrated window) recorded in the normal state (35 K) with 63 eV photons. Open circles and filled triangles correspond to $k_F$ locations of the $\gamma$ and $\kappa$ bands, respectively. (b) ARPES intensity plot ($h\nu=63$ eV) for a cut along the $\bar\Gamma$-$\bar{\textrm{M}}$ direction recorded at 35 K with a $p$ polarization. Guides to the eye are plotted for the various bands observed. (c) Same as (b) but using $s$-polarized photons. (d)-(e) EDCs corresponding to the cuts in (b) and (c), respectively. (f) ARPES intensity plot in the $k_z$-$k_x$ plane. The red and blue dashed lines indicate the $k_F$ locations. Reprinted with permission from \cite{XP_WangEPL2012}, copyright \copyright\xspace (2012) by the European Physical Society.}
\end{figure}

Despite its initial qualitative success in describing the SC properties of the Fe-based superconductors, the faith in the quasi-nesting scenario was not to last. The first major argument against this model is based on the discovery of superconductivity in the 122-ferrochalcogenides A$_x$Fe$_{2-x}$Se$_2$ \cite{JG_Guo_PRB2010,FangMH_EPL2011}, which have the same basic crystal structure as the 122-ferropnictide systems, as well as similarly high $T_c$ values. In comparison to the 122-ferropnictides, these systems are heavily-electron-doped. Consequently, large electron-like pockets are observed by ARPES at the M point, as shown in Fig. \ref{ch122FS}a. More significantly, their FS topology is exempt of any hole-like FS pocket, which prevents electron-hole quasi-nesting \cite{Qian_PRL2011,XP_WangEPL2011,D_MouPRL2011, Y_Zhang_NatureMat2011,XP_WangEPL2012}. In contrast, Fig. \ref{ch122FS} indicates a small 3D pocket ($\kappa$) observed at Z$(0,0,\pi/c')$ \cite{Y_Zhang_NatureMat2011,ZH_LiuPRL109}, which derives mainly from the Se $4p_z$ orbital, as deduced from polarization and photon energy dependent measurements \cite{ZH_LiuPRL109}. Indeed, the $\kappa$ band is detected only in configurations for which there is a finite component $A_z$ of the light polarization perpendicular to the sample surface. For example, Figs. \ref{ch122FS}b and \ref{ch122FS}d show clearly the $\kappa$ band on data recorded at the Swiss Light Source using $\pi$ (or $p$) configuration \cite{XP_WangEPL2012}, with a non-zero $A_z$ component. This band is not observed in pure $\sigma$ (or $s$) polarization, as illustrated in Figs. \ref{ch122FS}c and \ref{ch122FS}e. The situation is reversed when using pure $\pi$ polarization and $\sigma+A_z$ polarizations, such as at the Synchrotron Radiation Center \cite{ZH_LiuPRL109}. 

Even without analysing the SC gap structure, the observation of high-$T_c$ superconductivity in the absence of hole-like FS pocket is a strong and direct evidence against the quasi-nesting model, at least for the 122-ferrochalcogenides. Actually, the consequences to the pairing mechanism go much beyond and place \emph{all} the FS-driven pairing mechanisms into serious dilemmas. The huge price to pay for continuing to support the idea that electron-hole quasi-nesting mainly controls the pairing of electrons in the ferropnictide superconductors and in the 11-ferrochalcogenide superconductors is to admit the existence of a different, and yet still unconventional, pairing mechanism in the 122-ferrochalcogenides. For example, one could assume different intra-pocket and inter-pocket scattering parameters \cite{KhodasPRL108}. Even though mathematical solutions to this problem can be obtained, the physical justification for strong modifications of these parameters from one compound to another is not easy. 

An alternative scenario in which a FS-driven pairing mechanism would prevail would consist in saying that the pairing mechanism is controlled by the M-centred electron-like FS pockets. In this case, the presence or absence of $\Gamma$-centred hole-like FS pockets would not be critical to the superconductivity of the Fe-based superconductors. However, this assumption would be contradictory with the observation of a larger gap size on hole-like FSs than electron-like FSs in some materials, like BaCo$_{1.85}$Fe$_{0.15}$As$_2$, where a 7 meV SC gap is reported on the hole-like $\beta$ band, in contrast to a 4.5 meV gap on the electron-like FSs \cite{Terashima_PNAS2009}. More importantly, the assumption that only the electron-like FSs are important is in contradiction with the observation of Fe-based superconductivity at 9 K without electron-like FS pocket and with large $2\Delta/k_BT_c$ ratios in Ba$_{0.1}$K$_{0.9}$Fe$_2$As$_2$ \cite{Nan_XuPRB88}. Indeed, Xu $\emph{et al.}$ \cite{Nan_XuPRB88} showed that the Ba$_{1-x}$K$_x$Fe$_2$As$_2$ system encounters a Lifshitz transition \cite{Lifshitz_JETP11} (in fact there should be a series of Lifshitz transitions) between $x=0.7$ and $x=0.9$, by which the electron-like FSs pockets at M are replaced by the small hole-like $\varepsilon$ FSs pockets characterizing KFe$_2$As$_2$ \cite{Sato_PRL2009,Yoshida_JCPS72}. Because it involves only a shift of the chemical potential, this Lifshitz transition is fundamentally different from the one reported in the Ba(Fe$_{1-x}$Co$_x$)$_2$As$_2$ family \cite{C_LiuNaturePhys2010}, in which the small pockets attributed to a Dirac cone in the parent compound BaFe$_2$As$_2$ \cite{RichardPRL2010,Ran_PRB79,Harrison_PRB80} disappear after the suppression of the long-range AF order that follows Co doping. In the framework of the weak coupling approaches, this Lifshitz transition should have a dramatic impact on superconductivity since it destroys all possibilities of electron-hole quasi-nesting. However, albeit for a small slope change around $x=0.8$, there is no sudden drop of $T_c(x)$ in the phase diagram of Ba$_{1-x}$K$_x$Fe$_2$As$_2$ \cite{Rotter_Angew2008}, and the system evolves smoothly from the emergence of superconductivity in the under-doped regime up to $x=1$, suggesting that the same pairing mechanism is responsible for unconventional superconductivity in this family of materials. Consequently, neither the hole-like FSs at $\Gamma$ nor the electron-like FSs at M seem to be essential for Fe-based superconductivity. In addition, this approach focussing only on the electron-like FSs lacks of a fundamental physical support. Indeed, the FS topology of the 122-ferrochalcogenides is not ``special", in the sense that it does not emphasize on any particular wave vector, in opposition to the AF wave vector in the ferropnictides. 

An important issue to be discussed at this point is the inhomogeneities in the 122-ferropnictides. In fact, these materials are widely believed to show phase separation. Some of the most striking evidences include works on tunneling electron microscopy (TEM) \cite{YJ_Yan_SciRep2,ZW_WangJPC116,Z_WangPRB83}, scanning electron microscopy (SEM) \cite{ZW_WangJPC116,Speller_SST25,X_DingNCOMM2013,Y_LiuPRB86}, scanning tunneling microscopy (STM) \cite{W_LiNPHYS2011,W_LiPRL109,X_DingNCOMM2013,P_Cai_PRB85}, nuclear magnetic resonance (NMR) \cite{TexierPRL108}, M\"{o}ssbauer spectroscopy \cite{KsenofontovPRB84}, scanning nanofocussed x-ray diffraction \cite{RicciPRB84}, and near-field optical microscopy and low-energy muon spin rotation \cite{CharnukhaPRL109}. Although it is not a space-resolved probe, ARPES data also suggest phase separation. Using different samples and different cleaves of the same samples, one study showed the existence of different electronic structures ranging from metallic to Mott-insulating \cite{F_ChenPRB81}. Consistently, the presence of a large incoherent peak about 0.8 eV below $E_F$, for which the temperature evolution shows a metal to insulator crossover at a temperature coinciding with a hump in the resistivity data, had been interpreted previously as a signature of Mott physics \cite{XP_WangEPL2011}. 

Having exposed the occurrence of phase separation in the 122-ferrochacogenides, one could argue that the FS topology determined by ARPES for these materials is not representative of the SC phase, and perhaps of the Fe-based superconductors. However, apart for some minor relative band shifts and a chemical potential shift, the electronic structure is really consistent with what one would expect for a Fe-based superconductor, with hole-like bands centred at the $\Gamma$ point, here band-gapped by about 40-50 meV, and electron-like bands centred at the M point, as illustrated in Fig. \ref{ch122FS}. A qualitative agreement is also found with LDA band calculations \cite{Qian_PRL2011}. Therefore, it is very difficult to question the authenticity of the observed electronic structure as coming from a Fe-Se layer. As we will discuss below, it is precisely on this particular electronic band structure that a large SC gap, closing at the bulk $T_c$, is detected \cite{XP_WangEPL2011,D_MouPRL2011, Y_Zhang_NatureMat2011,XP_WangEPL2012}. This confirms that high-$T_c$ superconductivity without hole-like FS pockets can exist in a Fe-based superconductor. 

Results similar to the 122-ferrochalcogenides have also been reported on single-layer films of FeSe with $T_c$'s exceeding 55 K \cite{D_LiuNCOMM2012,S_TanNMAT12,S_HeNMAT2013,PengPRL112}. Since the films are annealed in vacuum, it is not clear whether the stoichiometry of the films or of the oxide substrates on which these films are grown, is changed during the process. Nevertheless, the electronic band structure corresponds to that of a Fe-based superconductor and as with the 122-ferrochalcogenides, large SC gaps are observed, supporting the claim that electron-hole quasi-nesting is not essential for Fe-based superconductivity. 

\begin{figure}[!t]
\begin{center}
\includegraphics[width=15 cm]{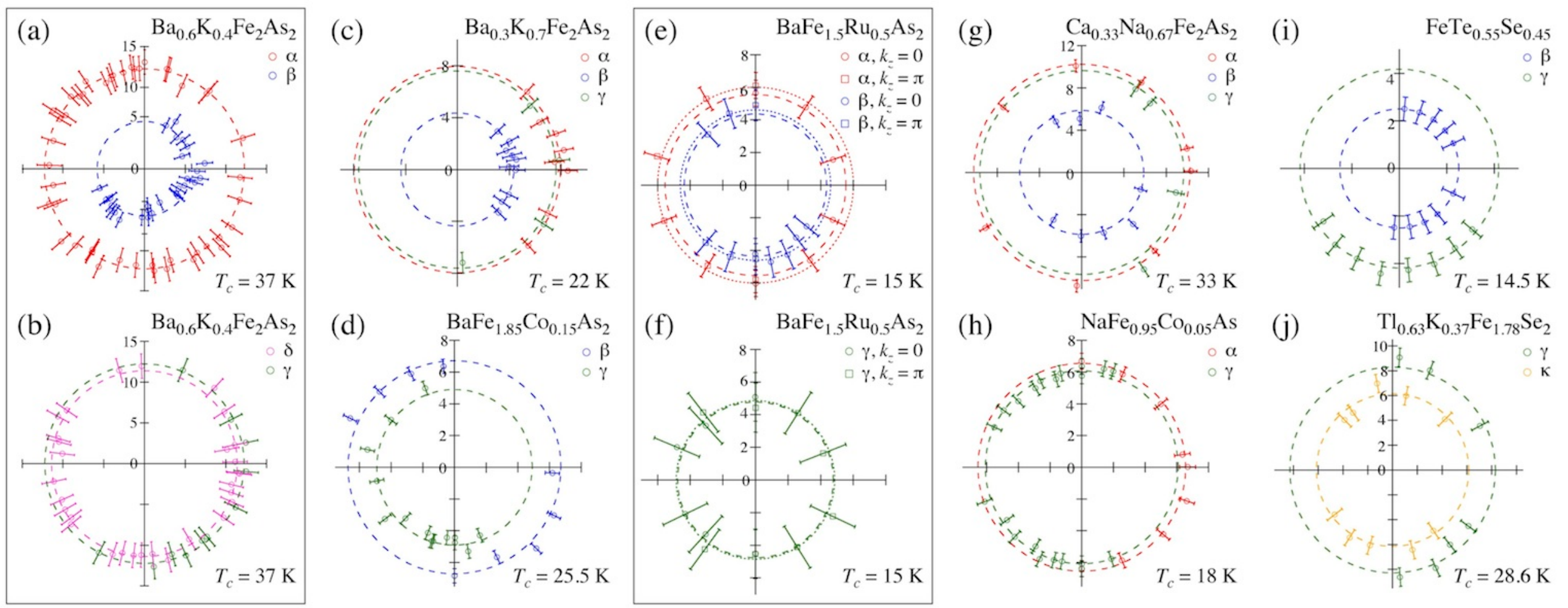}
\end{center}
\caption{\label{Fig_polar}(Colour online). Polar representations of the SC gap of several Fe-based superconductors. The polar angle is defined around the $\Gamma$ and M points, with 0 corresponding to the $\Gamma$-M high symmetry line. The large circles correspond to the average SC gaps. (a), (b) Ba$_{0.6}$K$_{0.4}$Fe$_2$As$_2$, data extracted from Ref. \cite{Nakayama_EPL2009}; (c) Ba$_{0.3}$K$_{0.7}$Fe$_2$As$_2$, data extracted from Ref. \cite{Nakayama_PRB2011}; (d) BaFe$_{1.85}$Co$_{0.15}$As$_2$, data extracted from Ref. \cite{Terashima_PNAS2009}; (e), (f) BaFe$_{1.5}$Ru$_{0.5}$As$_2$, data extracted from Ref. \cite{Nan_XuPRB87}, the dashed and dotted circles correspond to the average data recorded at $k_z\sim 0$ and $k_z\sim\pi$, respectively; (g) Ca$_{0.33}$Na$_{0.67}$Fe$_2$As$_2$, data extracted from Ref. \cite{YB_Shi_CPL31}; (h) Na$_{0.3}$Fe$_{0.95}$Co$_{0.05}$As, data extracted from Ref. \cite{ZH_LiuPRB84}; (i) FeTe$_{0.55}$Se$_{0.45}$, data extracted from Ref. \cite{H_Miao_PRB2012}; (j) Tl$_{0.63}$K$_{0.37}$Fe$_{1.78}$Se$_2$, data extracted from Ref. \cite{XP_WangEPL2012}.}
\end{figure}

In addition to the FS topology, ARPES measurements of the SC gap are also inconsistent with a FS-driven pairing mechanism. Low-energy interactions are very sensitive to the size and shape of the different FS pockets, and thus weak coupling approaches strongly suggest that nodes or strong gap anisotropies should be observed. In Fig. \ref{Fig_polar} we display a series of polar plots representing the SC gap amplitude around the $\Gamma$ ($\alpha$, $\beta$ and $\kappa$ bands) and M ($\gamma$ and $\delta$ bands) points. Not only the amplitude of the SC gap varies from one compound to the other, it varies also from one FS pocket to the other. However, all the examples shown in Fig. \ref{Fig_polar} show SC gaps that are more or less isotropic. This is even true for the 122-ferrochalcogenides, as shown in Fig. \ref{Fig_polar}(j), for which most weak coupling approaches predicted a $d$-wave SC gap \cite{WangFa_EPL2011,Mazin_PRB84,Maier_PRB83,Saito_PRB83,Das_PRB2011}. The absence of noticeable anisotropy, even on the small 3D $\kappa$ pocket \cite{Y_Zhang_NatureMat2011,XP_WangEPL2012}, clearly invalidates these scenarios. Although this leaves the weak-coupling approaches with a \emph{paradox} \cite{Hirschfeld_RoPP2011}, it is also remarkable that this observation of nodeless SC gaps is done for a variety of crystal structures, with different cleaved surfaces exposed for the ARPES measurements, thus reinforcing the reliability of the ARPES data. In fact, only a few systems deviate from this general pattern, and they will be the subject of Chapter \ref{section_nodes}. 

\section{The strong coupling approach}
\label{section_strong_coupling}

The failure of the FS-driven pairing mechanisms calls for alternative explanations of Fe-based superconductivity. The extreme opposite to the weak coupling approaches are the strong coupling approaches, in which the pairing of electrons comes from short-range interactions. In other words, the pairing process is better defined in the real space, and thus the FS topology does not play a critical role in the pairing itself. Of course, one can define a whole series of theories for couplings between the weak and strong coupling limits, which are usually refered to as intermediate coupling. To simplify the current discussion though, here we extend the terminology of strong coupling to englobe all theories for which the pairing is not specifically FS-driven. 

There are several physical justifications to the importance of short-range interactions and to the strong coupling approach in the high-$T_c$ materials. Unlike conventional superconductors, the Fe-based superconductors and the cuprates are particularly resistant to disorder. In the Ba$_{1-x}$K$_x$Fe$_2$As$_2$ family, for example, the highest $T_c$ occurs for $x=0.4$, which corresponds to an intrinsically disordered (Ba,K) layer. Even more surprisingly, an in-plane doping as large as 15 \% is necessary to optimize the $T_c$ of BaFe$_{2-x}$Co$_{x}$As$_2$. This suggests that the size of the Cooper pairs, and thus the range of the interactions causing their formation, must be relatively small. This is confirmed by the observation of large critical magnetic fields $H_{c2}$ in these materials \cite{GurevichRoPP74}. 

The amplitude of the SC gap provides another justification consistent with the strong coupling limit. As we can deduce from Fig. \ref{Fig_polar}, $2\Delta/k_BT_c$ ratios larger than the 3.5 BCS ratio are commonly observed in the Fe-based materials, even by a factor of 2. For example, a ratio of 7.5 was reported in Ba$_{0.6}$K$_{0.4}$Fe$_2$As$_2$ for the hole-like $\alpha$ FS and the electron-like $\gamma$ FS \cite{Ding_EPL}, a ratio that remains more or less constant as $T_c$ drops to 26 K upon underdoping \cite{YM_Xu_UD} or to 22 K upon overdoping \cite{Nakayama_PRB2011}. This ratio becomes 6.8 in Ca$_{0.33}$Na$_{0.67}$Fe$_2$As$_2$ \cite{YB_Shi_CPL31} and despite a much smaller $T_c$ of 15 K, a ratio of 9 was even reported for the $\alpha$ band in BaFe$_{1.5}$Ru$_{0.5}$As$_2$ \cite{Nan_XuPRB87}. Such large $2\Delta/k_BT_c$ ratios are not unique to the 122-ferropnictides. A ratio of 8 was obtained on the 111-ferropnictide Na$_{0.3}$Fe$_{0.95}$Co$_{0.05}$As \cite{ZH_LiuPRB84}, whereas ratios as high as 6.7 and 7 where observed in the 11-ferrochalcogenide FeTe$_{0.55}$Se$_{0.45}$ \cite{H_Miao_PRB2012} and in the 122-ferrochalcogenide Tl$_{0.63}$K$_{0.37}$Fe$_{1.78}$Se$_2$ \cite{XP_WangEPL2011}, respectively. As for the monolayer FeSe thin films, which exhibit the largest SC gaps and the highest $T_c$'s, typical ratios of 6-7 are extracted from the SC gap measurements \cite{D_LiuNCOMM2012,S_TanNMAT12,S_HeNMAT2013,PengPRL112}.

Even though we concluded above that the pairing interaction in the strong coupling limit is better defined in the real space, short-range interactions have a direct impact on the properties that are measured in the momentum space. As a first approximation, these short-range interactions between a site $i$ and a site $j$, like the ones that would induce the electron pairing, can be considered as proportional to $\delta(\vec{r}_{ij})$ functions, with $\vec{r}_{ij}$ representing the vector position between the two sites. Naturally, the Fourier transforms of such functions lead to simple combinations of sine and cosine functions defined all over the momentum space. Ref. \cite{HuJP_SR2012} provides a list of such functions for interactions between the first, the second and the third nearest neighbours. Assuming the validity of the strong coupling approach, the SC gap function is thus represented in the momentum space as a \emph{global} function $\Delta(\vec{k})$ that is \emph{a priori} irrelevant of the relative locations of the various FS sheets. 

Considering the proximity between the AF and SC states in the phase diagram of the Fe-based superconductors, it is natural the consider the AF interactions as a potential glue for the SC pairing. Due to the strong bi-dimensional (2D) character of the crystal structure, strong fluctuations of the in-plane interactions survive to the collapse of long-range 3D AF ordering. To estimate their strength, one can investigate the AF ordering of the parent compounds. Indeed, their magnetic ordering can be characterized by the $J_1-J_2-J_3$ model, where $J_1$, $J_2$ and $J_3$ represent the strength of the exchange interactions between the first, second and third neighbours, respectively. The values of $J_1$, $J_2$ and $J_3$ can be derived experimentally by parameterising the spin wave dispersions measured by inelastic neutron scattering on the parent compounds. The results indicate clearly which parameters are the most important for the magnetic ordering, and thus indirectly to the electron pairing. Keeping only the relevant AF parameters, one can thus easily construct the proper global SC gap function, even before any ARPES SC gap measurement is made.  

\begin{figure}[!t]
\begin{center}
\includegraphics[width=15 cm]{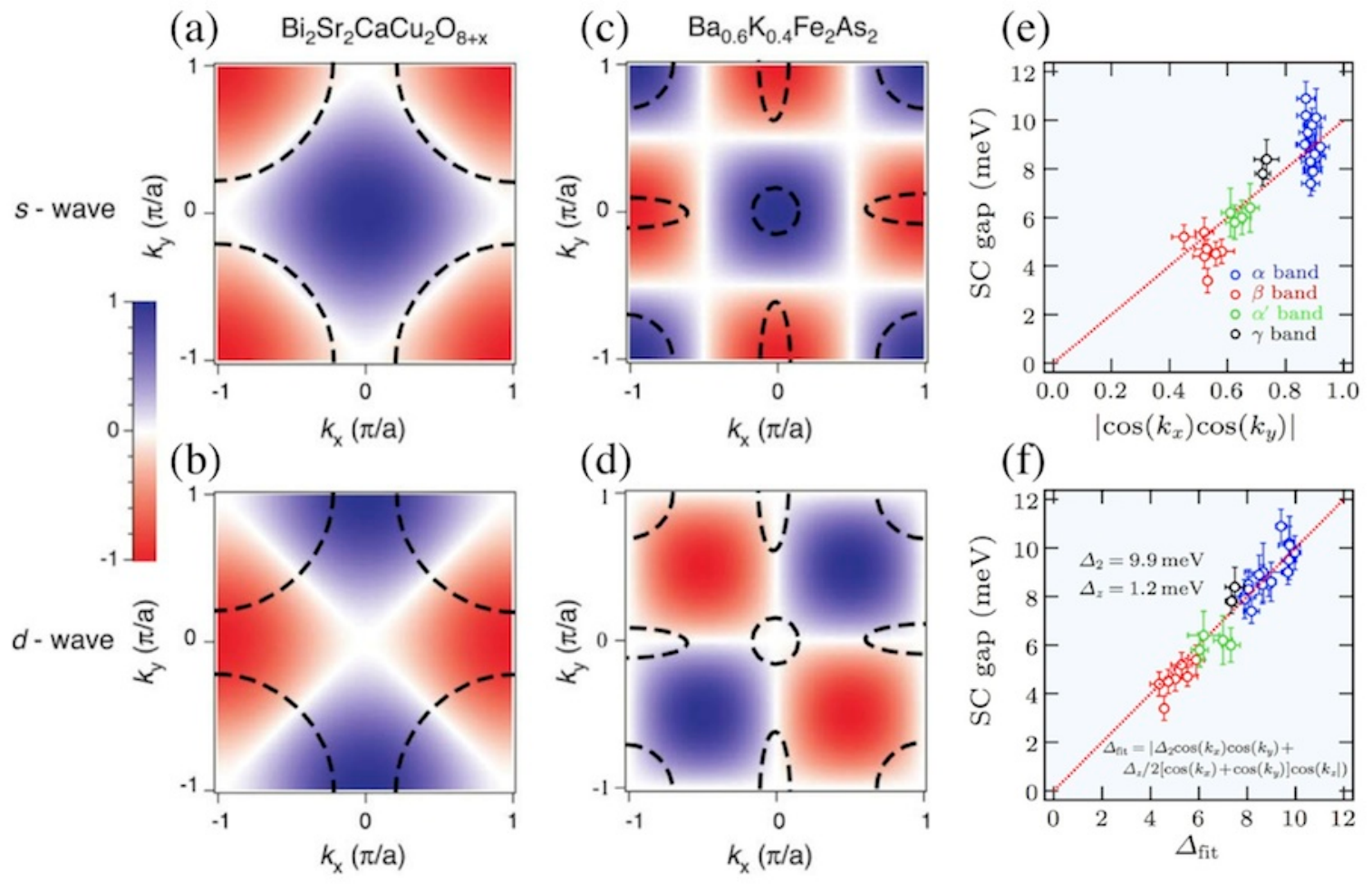}
\end{center}
\caption{\label{gapfunc}(Colour online). (a)-(d) Visualization of the overlap between FS and gap functions: (a) $s$-wave $\cos(k_x)+\cos(k_y)$ for optimally doped cuprate Bi$_2$Sr$_2$CaCu$_2$O$_{8+x}$; (b) $d$-wave $\cos(k_x)-\cos(k_y)$ for Bi$_2$Sr$_2$CaCu$_2$O$_{8+x}$. (c) $s$-wave $\cos(k_x)\cos(k_y)$ for optimally doped ferropnictide Ba$_{0.6}$K$_{0.4}$Fe$_2$As$_2$. (d) $d$-wave $\sin(k_x)\sin(k_y)$ for Ba$_{0.6}$K$_{0.4}$Fe$_2$As$_2$. The colour bar indicates the values of the SC order parameters. (e) SC gap magnitude on the various FSs of Ca$_{0.33}$Na$_{0.67}$Fe$_2$As$_2$ as a function of the global gap function $|\cos(k_x)\cos(k_y)|$. (f) The same as (e) but for the gap function $\Delta_{fit} = |\Delta_2\cos(k_x)\cos(ky) + (\frac{1}{2}\Delta_z)[\cos(k_x)+\cos(k_y)]\cos(k_z)|$. Panels (a)-(d) are reprinted by permission from Macmillan Publishers Ltd: Sci. Rep. \cite{HuJP_SR2012}, copyright \copyright\xspace (2012). Panels (e) and (f) are reprinted with permission from \cite{YB_Shi_CPL31}, copyright \copyright\xspace (2014) by IOP publishing.}
\end{figure}

Let's illustrate this procedure by considering the case of the cuprate superconductors, where $J_1$ is the dominant AF exchange parameter. In this case, we naturally derive a $s$-wave SC gap function of the form $\Delta(\vec{k})=\frac{1}{2}\Delta_1[\cos(k_x)+\cos(k_y)]$ and a $d$-wave SC gap function of the form $\Delta(\vec{k})=\frac{1}{2}\Delta_1[\cos(k_x)-\cos(k_y)]$ \cite{HuJP_SR2012}. The amplitudes of these functions are represented by a colour scale in Figs. \ref{gapfunc}a and \ref{gapfunc}b, respectively. Although the FS does not play a direct role in the pairing of electrons within the strong coupling approach, it is somewhat indirectly important to distinguish which of these possible gap functions is the most favourable. Obviously, the most favourable one would be the one lowering the most the total energy by gapping electronic states at the FS. In other words, the proper SC gap function is the one that has the best overlap with the FS, which can also be quantified \cite{HuJP_SR2012}. Keeping in mind that for minimizing energy we need considering the square of the gap function, we can see from Fig. \ref{gapfunc}b, that the overlap between the $J_1$-dominant $s$-wave gap function and the typical FS of a cuprate superconductor is small. Although the same FS crosses regions of zero amplitude when overlapped with the $J_1$-dominant $d$-wave gap function, giving rise to nodes in the gap function, a strong overlap is observed near $(\pi,0)$, commonly referred to as the antinodal region. Consequently, the strong coupling approach necessarily favours a $d$-wave SC gap in the cuprates.

The situation is different in the ferropnictides. For these materials, inelastic neutron scattering experiments indicate a ferromagnetic $J_1$ parameter, which is not involved in the pairing mechanism. The dominant AF exchange constant is thus $J_2$. Therefore, the corresponding $s$-wave and $d$-wave gap functions that are naturally derived differ from that in the cuprates. These functions, which correspond to $\Delta(\vec{k})=\Delta_2\cos(k_x)\cos(k_y)$ and $\Delta(\vec{k})=\Delta_2\sin(k_x)\sin(k_y)$, are illustrated in Figs. \ref{gapfunc}c and \ref{gapfunc}d, respectively. As with the previous analysis on the cuprates, we overlap on these figures the typical FS of a ferropnictide. While the overlap is quite poor for the $d$-wave function, the overlap with the $s$-wave function is pretty good, suggesting that in the ferropnictide a $s$-wave pairing symmetry prevails. Assuming that the phase of the SC gap is fixed by this one-parameter global gap function, an assumption that will be discussed in further details in the next section, the strong coupling approach thus naturally reproduces the $s_{\pm}$ SC gap function, with an opposite phase sign for the $\Gamma$-centred hole-like FSs and the M-centred electron-like FSs. 

For FS pockets that are more or less circular and that are centred either at the $\Gamma$ or at the M point, the model presented here leads to more or less isotropic SC gaps, with an amplitude decreasing with the FS size. This situation is well illustrated with the 122-ferropnictide Ca$_{0.33}$Na$_{0.67}$Fe$_2$As$_2$, which has FS pockets of different sizes \cite{YB_Shi_CPL31}. In Fig. \ref{gapfunc}e we plot the SC gap amplitude in this material as a function of $|\cos(k_x)\cos(k_y)|$. Considering that only one global parameter is used in the fit for all the FSs, the agreement is pretty good. The largest gap, $\Delta_{\alpha}$, is found along the $\alpha$ band, which has the smallest FS, as shown in Fig. \ref{Fig_polar}g. Then come respectively $\Delta_{\gamma}$, $\Delta_{\alpha'}$ and $\Delta_{\beta}$, corresponding to the gap amplitude along the $\gamma$, $\alpha'$ (observed at $k_z=\pi$) and $\beta$ bands. This ranking is exactly the same as for the FS sizes. To improve the fit even further, it is necessary to consider the dispersion along $k_z$. Indeed, some bands in the Fe-based superconductors exhibit a non-negligible 3D character. As first shown for Ba$_{0.6}$K$_{0.4}$Fe$_2$As$_2$ \cite{Y_Zhang_PRL2010,YM_Xu_NPhys2011}, this affects the SC gap amplitude as well, which varies slightly along $k_z$. Following that study and similar works on the 122-ferropnictides BaFe$_2$(As$_{0.7}$P$_{0.3}$)$_2$ \cite{Y_Zhang_NaturePhys2012} and Ba(Fe$_{0.75}$Ru$_{0.25}$)$_2$As$_2$ \cite{Nan_XuPRB87}, one SC gap function can be modified to include an inter-layer coupling term:

\begin{equation}
|\Delta(\vec{k})|=|\Delta_2\cos(k_x)\cos(k_y)+\frac{1}{2}\Delta_z[\cos(k_x)+\cos(k_y)]\cos(k_z)|.
\end{equation}  

The fit of the experimental data to that previous gap function, illustrated in Fig. \ref{gapfunc}f, leads to $\Delta_2=9.9$ meV and $\Delta_z=1.2$ meV \cite{YB_Shi_CPL31}. Interestingly, the $\Delta_2/\Delta_z=8.3$ ratio is similar to the $J_2/J_z=7$ ratio determined by inelastic neutron scattering on the parent compound CaFe$_2$As$_2$ \cite{McQueeneyPRL2008}. Because the data spread over a wide range of the gap function, the pretty good agreement of the fit with the experimental data, despite the use of only 2 global parameters, is a clear indication that the SC gap amplitude is, at least at the first order, a function of the absolute position in the momentum space, and is thus independent of the FS topology as well as the intra-band and inter-band scattering interactions, the latter interactions being functions of the momentum-transfer. It is also worth mentioning that despite much weakened quasi-nesting conditions in Ca$_{1-x}$Na$_{x}$Fe$_2$As$_2$ \cite{YB_Shi_CPL31,Evtushinsky_PRB87} as compared to Ba$_{0.6}$K$_{0.4}$Fe$_2$As$_2$, the $T_c$ values and the SC gap amplitudes are comparable, which contradicts the quasi-nesting model. 

The observation of SC gap amplitude depending on the FS size can also be made for other pnictide compounds. As illustrated in Fig. \ref{quasi_nesting}a, the size of the $\alpha$, $\gamma$ and $\delta$ FS pockets is almost the same, and thus the SC gap amplitudes measured along these 3 FSs are very similar. Clearly though, the $\delta$ FS encloses completely the $\gamma$ FS, and its size is thus larger. Consistently with the strong coupling approach, the amplitude of the SC gap along the former FS is slightly smaller than that observed on the latter one, as shown in Fig. \ref{Fig_polar}b. Within this framework, it is also easy to understand why the $\beta$ FS, much larger than all the other ones, carries a much smaller SC gap. It is also without any surprise that the corresponding $2\Delta_{\beta}/k_BT_c$, which suggests a weak coupling in Ba$_{0.6}$K$_{0.4}$Fe$_2$As$_2$ \cite{Ding_EPL,L_Zhao,Nakayama_EPL2009}, switches to a strong coupling value after the $\beta$ FS shrinks significantly in BaFe$_{1.85}$Co$_{0.15}$As$_2$ due to electron-doping \cite{Terashima_PNAS2009}. A similar global function can also describe the SC gap in the 111-ferropnictide NaFe$_{0.95}$Co$_{0.05}$As \cite{ZH_LiuPRB84}. In this material, the $\Gamma$-centred hole-like FS pocket is smaller than the M-centred electron-like FS pockets. Accordingly, the average SC gap size is larger along the former one. Moreover, independent fits of the SC gaps along both FSs lead to very similar global gap parameters (6.8 vs 6.5 meV) \cite{ZH_LiuPRB84}, indicating that a single global parameter is sufficient to describe the SC gap on both FSs. 

The situation in the ferrochalcogenides differs slightly from that of the ferropnictides and offers a critical test for the validity of the strong coupling approach. Unlike the ferropnictides, for which the magnetic ordering is sufficiently well described by considering only exchange interactions up to the next-nearest neighbours, inelastic neutron scattering experiments indicate that one needs to consider the exchange parameter $J_3$ as well, which corresponds to interactions between the next-next-nearest neighbours. As a consequence, the SC gap function illustrated in Fig. \ref{gapfunc}c is insufficient to characterize the SC gap in these materials. The appropriate additional SC gap function corresponding to a $s$-wave supported by the parameter $J_3$ takes the form $\frac{1}{2}\Delta_3(\cos(2k_x)+\cos(2k_y))$ \cite{HuJP_SR2012}, and thus the global gap function expected to describe the ferrochalcogenides is:

\begin{equation}
\label{eq_J2_J3}
|\Delta(\vec{k})|=|\Delta_2\cos(k_x)\cos(k_y)-\frac{1}{2}\Delta_3[\cos(2k_x)+\cos(2k_y)]|.
\end{equation}   

Assuming that $|J_2|>|J_3|$ and thus that $|\Delta_2|>|\Delta_3|$, the main effect of the introduction of the new term in $\Delta_3$ is to induce an asymmetry in the SC gap function between the SC gaps at the $\Gamma$ and M points. While this asymmetry favours the $\Gamma$ point for $\Delta_3<0$, the opposite scenario occurs when $\Delta_3>0$. Interestingly, the SC gap data recorded on FeTe$_{0.55}$Se$_{0.45}$ are not well reproduced by the simple $s_{\pm}$ gap function. Indeed, although their FS size is comparable, the SC gap amplitude along the M-centred electron-like $\gamma$ FS pocket is almost twice as large as the one measured on the $\Gamma$-centred hole-like $\beta$ FS pocket \cite{H_Miao_PRB2012}. In contrast, a fit to Eq. (\ref{eq_J2_J3}) gives a good agreement with $\Delta_2=3.55$ meV and $\Delta_3=0.95$ meV. The ratio $\Delta_2/\Delta_3$ of these two global parameters is similar to the $J_2/J_3$ (22/7) ratio obtained from inelastic neutron scattering \cite{LipscombePRL2011}. Similar ARPES results have been reported for the 122-chalcogenide Tl$_{0.63}$K$_{0.37}$Fe$_{1.78}$Se$_2$. Despite a much smaller size, the small 3D electron-like $\kappa$ FS centred at the Z point carries a SC gap that is smaller than the one measured on the M-centred hole-like $\gamma$ FS pocket. In this case, a fit to Eq. (\ref{eq_J2_J3}) leads to $\Delta_2=9.7$ meV and $\Delta_3=3.4$ meV \cite{XP_WangEPL2012}. Because the global gap function indicates that the SC gaps around the M point are larger than those around the $\Gamma$ point, the presence or absence of hole-like FS pockets centred at $\Gamma$ is not essential for the stability of the SC state, thus reinforcing the statement that unlike the quasi-nesting model, the strong coupling approach remains valid in the 122-chalcogenides. 

The strong coupling approach has been confirmed further by recent ARPES observations on lightly Co-doped LiFe$_{1-x}$Co$_x$As \cite{H_Miao_NCOMM6}. With electron-doping, the $\alpha$ band is quickly band-gapped in this material. Nevertheless, the band structure at the $\Gamma$ point is modified below $T_c$, a modification that can still be described by the opening of a SC gap following Eq. (\ref{gap_BCS}) with a negative Fermi energy. The authors argued that this SC gap cannot be the result of a proximity effect since its magnitude is larger than that of the other FSs, which cannot be understood in terms of low-energy interactions.

\section{Determination of the phase of the superconducting gap}
\label{section_phase}

As explained in the previous sections, the strong coupling approach is consistent not only with the FS topology of the Fe-based superconductors, but with the amplitude of the SC gap measured by ultra-high resolution ARPES experiments as well. However, these measurements do not provide a complete understanding of the SC state since the phase needs to be determined and ARPES does not give direct access to the phase. In the cuprate superconductors, the most convincing determination of the sign of the phase on different lobes of the $d$-wave gap function has been revealed from three grain-boundary Josephson experiments \cite{Kirtley_Nature373,Tsuei_Nature387}. Unfortunately, the multi-band nature of the electronic structure of the Fe-based superconductors prevents the design of similar experiments, and other ways to access the phase of the SC gap are necessary. Currently, the two most popular models describing the phase of the SC gap are the so-called $s_{\pm}$ \cite{MazinPRL2008,KurokiPRL101,SeoPRL2008} and $s_{++}$ \cite{Kontani_PRL104} models. While in the former model the phase of the SC gap are opposite at the $\Gamma$ and M points, it is everywhere the same in the latter one. In this section, we show that the most likely scenario is different from $s_{\pm}$ and $s_{++}$.

Several theoretical studies have investigated how the density-of-states would look like inside the SC gap in the presence of impurities. \cite{ParishPRB78,ParkerPRB78,ChubukovPRB78,BangPRB78}. Due to inter-band scattering, the formation of in-gap states is closely related to the relative phase of the SC gap on each FS sheet, which can be used to conclude which of the $s_{\pm}$ and $s_{++}$ gap structures can better describe the Fe-based superconductors \cite{Y_WangPRB87}. Although, these studies mainly focus on the density-of-states, it is interesting to look at this problem using a momentum-resolved probe like ARPES. 

\begin{figure}[!t]
\begin{center}
\includegraphics[width=15 cm]{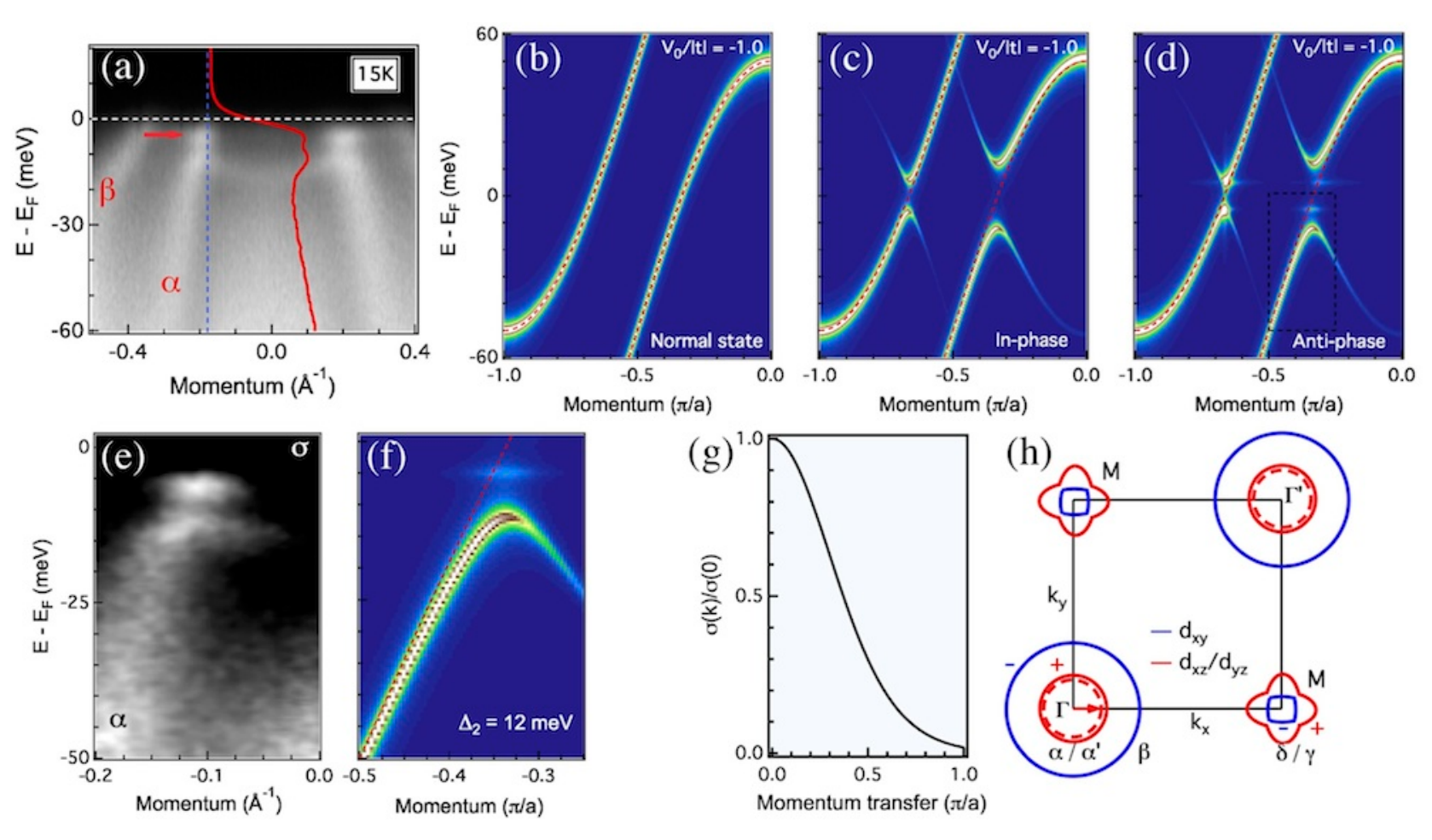}
\end{center}
\caption{\label{phase_paper}(Colour online). (a) ARPES intensity plot recorded at 15 K along the $\Gamma$-M high-symmetry line. The red arrows indicate an in-gap state. The EDC at $k_F$ is also displayed in red. (b)-(d) Numerical simulations of the spectral weight for a system with an impurity interacting with two dispersive bands according to Eq. [\ref{two_bands}] in the normal state ($\Delta_1=\Delta_2=0$), in the in-phase SC state ($\Delta_1=\frac{1}{2}\Delta_2\neq 0$), and in the anti-phase SC state ($\Delta_1=-\frac{1}{2}\Delta_2\neq 0$), respectively. In all cases, we used $V_0/|t|=-1$, where $|t|$ corresponds to half of the band width. The red dashed lines represent the electronic dispersion in the normal state. (e) Zoom on the impurity state found experimentally near the $\alpha$ band of Ba$_{0.6}$K$_{0.4}$Fe$_2$As$_2$. (f) Numerical simulation of the zoom near the impurity state corresponding to the dashed box from panel d (anti-phase SC). The red dashed line in f represents the bare band dispersion. (g) Calculated scattering strength as a function of momentum transfer. (h) Schematic FS of Ba$_{0.6}$K$_{0.4}$Fe$_2$As$_2$, with the FS sheets drawn in red and blue having opposite SC gap phase signs. Reprinted with permission from \cite{P_ZhangPRX2014}, copyright \copyright\xspace (2014) by the American Physical Society.}
\end{figure}

In Fig. \ref{phase_paper}a we show an ARPES intensity plot recored on Ba$_{0.6}$K$_{0.4}$Fe$_2$As$_2$, below $T_c$ \cite{P_ZhangPRX2014}. The electronic dispersion of the $\alpha$ band is clear. The band tops at 12 meV and then bends back towards the high energies, exactly as one would expect for a SC gap. Therefore, the $\alpha$ band exhibits a SC gap of 12 meV, as reported earlier \cite{Ding_EPL,L_Zhao, Nakayama_EPL2009}. Interestingly, an additional state is observed at lower energy, inside the SC gap, as indicated by a red arrow. This state found at 6 meV is dispersionless, and thus attributed to impurities. By using different light polarizations, Zhang \emph{et al.} showed that this state is also observed near the $k_F$ locations of the $\beta$ and $\alpha'$ bands, as well as near the electron-like bands at the M point, suggesting that the state involves scattering with all bands \cite{P_ZhangPRX2014}. Although the in-gap state does not correspond to a SC gap, it is observed only below $T_c$. In fact, the in-gap state feature has been associated with a SC gap with sub-BCS amplitude in a laser-ARPES study, whereas the 12 meV feature was assigned to a magnetic resonance mode or a coupling with orbital degrees of freedom \cite{Shimojima_Science332}. However, this interpretation is incompatible with the flatness of the in-gap feature and the observation of Bogoliubov dispersion at 12 meV. 

At the first sight, the observation of spectral intensity only near the various $k_F$ positions is a little counter-intuitive. However, it is what is expected for relatively weak scattering. This situation is also quite similar to that reported in Si-doped $\beta$-Ga$_2$O$_3$, a large gap semiconductor for which the ARPES spectra show a momentum space confinement $\Delta k$ \cite{RichardAPL2012} matching the real space confinement $\Delta r$ determined by STM \cite{IwayaAPL2011} through a $\Delta k\Delta r\approx 1$ relationship. Assuming that no bounded in-gap state should occur for a one-band $s$-wave system in the presence of non-magnetic impurities \cite{PW_AndersonJPCS11}, and in accord with the absence of report on magnetic impurities in as-grown (Ba,K)Fe$_2$As$_2$, Zhang \emph{et al.} simulated a two-band system with non-magnetic impurities with the Hamiltonian:

\begin{equation*}
H=\sum_{\vec{k},m,\sigma}\varepsilon_m(\vec{k})c^{\dagger}_{m,\vec{k},\sigma}c_{m,\vec{k},\sigma}\\
\end{equation*}
\begin{equation}
\label{two_bands}
+\sum_{\vec{k},m}\Delta_m(c^{\dagger}_{m,\vec{k},\uparrow}c^{\dagger}_{m,-\vec{k},\downarrow}+c_{m,\vec{k},\downarrow}c_{m,-\vec{k},\uparrow})+\frac{V_0}{2N}\sum_{m,n,\vec{k},\vec{k'},\sigma}c^{\dagger}_{m,\vec{k},\sigma}c_{n,\vec{k'},\sigma},
\end{equation}

\noindent where the first right hand side term represents the unperturbed Hamiltonian with the operator $c^{\dagger}_{\vec{k},\sigma}$($c_{\vec{k},\sigma}$) creating(anihilating) an electron of spin $\sigma$ and wave vector $\vec{k}$ and with $\varepsilon_m(\vec{k})$ describing the unperturbed electronic dispersion of band $m$. The second term accounts for superconductivity with a gap $\Delta_m$ on band $m$ while the third term defines scattering by $N$ punctual impurities characterized by an impurity potential $V_0$ \cite{BalatskyRMP}. The indexes $m$ and $n$ take the values 1 and 2 representing the two bands. The Hamiltonian above can be diagonalized numerically after dividing the first BZ into 500 points. More specifically, one can extract the spectral function $A(\vec{k},\omega)$ by using the equation:

\begin{equation}
\label{spectral_weight}
A(\vec{k},\omega)=-\frac{1}{\pi}\displaystyle\textrm{Im}\sum_m\frac{|\braket{\vec{k}}{m}|^2}{\omega-E_m+i\delta},
\end{equation} 

\noindent where the eigenvectors $\ket{m}$ with eigenvalues $E_m$ are projected into the momentum space. The resulting spectral functions obtained by using the diagonal terms of the Green's function are shown in Figs. \ref{phase_paper}(b)-\ref{phase_paper}(d) for three distinct cases. In Fig. \ref{phase_paper}(b) we show the simulation in the normal state, with $\Delta_1=\Delta_2=0$. Fig. \ref{phase_paper}(c) shows the simulation for the SC state with the SC gaps in-phase ($\Delta_1=\frac{1}{2}\Delta_2\neq 0$). The gap opening appears clearly, as well as the characteristic Bogoliubov dispersion. However, no extra feature is observed. This contrasts with the simulation given in Fig. \ref{phase_paper}(d), for which the two SC gaps are in anti-phase ($\Delta_1=-\frac{1}{2}\Delta_2\neq 0$). Indeed, in-gap impurity states are detected near the $k_F$ positions, above and below $E_F$. A zoom, displayed in Fig. \ref{phase_paper}(f), compares qualitatively pretty well with the experimental data shown in Fig. \ref{phase_paper}(e), suggesting that the phase of the SC gap is not constant in Ba$_{0.6}$K$_{0.4}$Fe$_2$As$_2$, thus ruling out the $s_{++}$ model \cite{P_ZhangPRX2014}. 

Further information can be obtained by investigating the nature of the scattering. As an improvement to a constant impurity potential $V_0$, one can consider the screened Coulomb potential:

\begin{equation}
\label{screening_r}V(r)=-\frac{e^2}{4\pi\epsilon_0 \sqrt{r^2+d^2}}e^{-\sqrt{r^2+d^2}/\lambda},
\end{equation}

\noindent where $d$ is the distance between the Fe and impurity planes, $r$ is the in-plane distance and $\lambda$ is the Thomas-Fermi screening length, which is estimated to be about 1 \AA\xspace for the $1.85\times 10^{21}$ cm$^{-3}$ electron density in Ba$_{0.55}$K$_{0.45}$Fe$_2$As$_2$ \cite{JohnstonAdv_Phys2010}. Zhang \emph{et al.} demonstrated that the trivial case of non-magnetic Ba$^{2+}$/K$^{+}$ disorder was the most likely to explain the experimental data on Ba$_{0.6}$K$_{0.4}$Fe$_2$As$_2$ \cite{P_ZhangPRX2014}. Their calculations of the scattering rate as a function of the momentum transfer, shown in Fig. \ref{phase_paper}(g), indicate that scattering with small momentum transfer, in other words between bands that are located closely in the momentum space, is largely favoured. Although the conventional $s_{\pm}$ cannot be ruled out, this latter observation suggests that the sign of the phase of one $\Gamma$-centred hole-like FS must differ from that of the two others, and that the two M-centred electron-like FSs carry opposite signs for the phase of the SC gap. One possible scenario coinciding with this rule, which is illustrated in Fig. \ref{phase_paper}(g), is the recently proposed anti-phase $s_{\pm}$ gap structure derived both from a four-site model \cite{X_LuPRB85} and first-principles calculations including the \emph{ab initio} determination of the two-particle vertex function \cite{Yin_ZP_orbital2014}. Both theoretical approaches found that the $d_{xy}$ and $d_{xz}/d_{yz}$ orbitals should exhibit opposite signs of the SC gap. In addition, a theoretical study suggested that the introduction of an odd-parity term \cite{HuJP_PRX3} can lead to the exact anti-phase $s_{\pm}$ state \cite{Hao_PRB89,HuJP_PRX2012}. 

It is important to note that weak coupling approaches have also been used to reproduce the sign change suggested by ARPES. Indeed, for systems without $\Gamma$-centred hole-like FS pocket, the odd parity pairing is the same as the bonding-antibonding $s_{\pm}$ suggested in Refs. \cite{KhodasPRL108,Mazin_PRB84}. Interestingly, a $s^h_{\pm}$ SC gap pattern, for which only the large $d_{xy}$ FS pocket is sign-reversed, has been derived from low-energy orbital fluctuations-driven superconductivity in the presence of weak spin fluctuations \cite{Saito_PRB90}. Although these mathematical solutions cannot be totally ruled out, the ARPES results on the FS topology and the SC gap amplitude confirm the robustness of the strong coupling approach in explaining Fe-based and Cu-based superconductivity in a more universal fashion \cite{HuJP_SR2012}.

\section{Nodes and superconducting gap anisotropy}
\label{section_nodes}

\begin{figure}[!t]
\begin{center}
\includegraphics[width=8 cm]{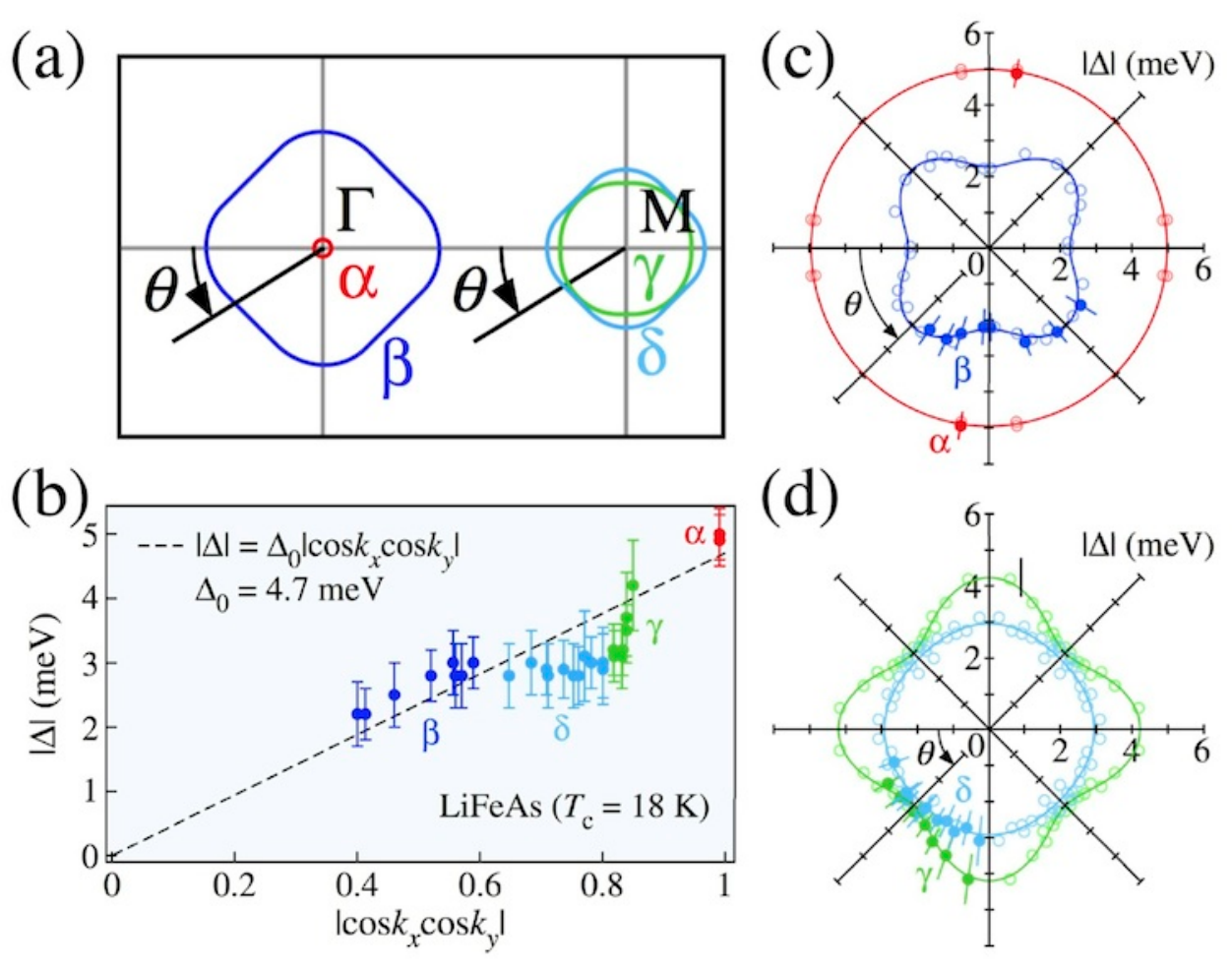}
\end{center}
\caption{\label{Fig_111}(Colour online). (a) Schematic FS of LiFeAs and definition of the FS angle $\theta$. (b) Plot of the SC gap size as a function of $|\cos(k_x)\cos(k_y)|$. The fitting result assuming the gap function $|\Delta|=|\Delta_2\cos(k_x)\cos(k_y)|$ is indicated by a black dashed line. (c) Polar plot of the SC gap size for the $\alpha$ and $\beta$ FSs as a function of the angle $\theta$ defined in panel (a). (d) Same as panel (c) but for the $\gamma$ and $\delta$ FSs. Filled circles in (c) and (d) are the original data, and open circles are the folded data, which take into account the fourfold symmetry. Solid curves show the fitting
results with $\Delta(\theta)=\bar{\Delta}_0+\bar{\Delta}_1\cos[4(\theta+\phi)]$. Reprinted with permission from \cite{UmezawaPRL2012}, copyright \copyright\xspace (2012) by the American Physical Society.}
\end{figure}

In section \ref{section_weak_coupling} we presented the observation of nodeless and isotropic SC gaps in most Fe-based superconductors as an evidence against FS-driven pairing mechanisms. However, a few ARPES studies suggest the presence of anisotropic SC gaps, and even SC gap nodes. The most obvious an undisputed case is the one of LiFeAs. Unlike most Fe-based superconductors, this LiFeAs is free of non-stoichiometric defects. In addition, the cleaved surface is non-polar and perfectly suited for high-resolution ARPES measurements. As shown in Fig. \ref{Fig_111}a, the FS of this material is characterized by weak electron-hole quasi-nesting conditions and by M-centred electron-like FS pockets with small eccentricity as compared to the (Ba,K)Fe$_2$As$_2$ family. 

The polar representations of the SC gap along the $\Gamma$-centred hole-like FSs and M-centred electron-like FSs are displayed in Figs. \ref{Fig_111}c and \ref{Fig_111}d, respectively. An anisotropic profile is observed for the $\beta$ band that cannot be neglected \cite{UmezawaPRL2012,BorisenkoSym2012}, the gap amplitude varying from about 2 meV along the $\Gamma$-M direction to 3 meV along $\Gamma$-X. As shown in Ref. \cite{Y_Huang_AIP2012}, this result is also perfectly consistent with STM data on the same material \cite{Allan_Science336}. The gap shape can clearly be fitted by a $\bar{\Delta}_0+\bar{\Delta}_1\cos[4(\theta+\phi)]$ function. Interestingly, the SC gap along the $\gamma$ FS also shows a strong anisotropy. In this case, a maximum gap of about 4 meV is found along $\Gamma$-M, in contrast to a 3 meV gap observed at 45$^{\circ}$ degrees from that direction. Obviously, the same kind of gap function can be used to fit the gap amplitude on the $\gamma$ FS. Analysing these oscillations, Borisenko \emph{et al.} concluded that low-energy orbital fluctuations assisted by phonons is the best explanation for superconductivity in LiFeAs \cite{BorisenkoSym2012}. The same results have been reanalysed by considering low-energy spin fluctuations as well, which can lead to a complex evolution of the order parameter from $s_{++}$ to $s^{h}_{\pm}$ \cite{Saito_PRB90}. 

The previous interpretation violates the rules of universality derived in the previous sections, from which we concluded that Fe-based superconductivity was not driven by the FS topology, and therefore could not be associated with low-energy fluctuations, whether coming from the spin or the orbital degrees of freedom. Yet, the observation of strong modulations of the SC gap around some FSs differs from the results obtained on most Fe-based superconductors, as illustrated in Fig. \ref{Fig_polar}, and requires an explanation from the strong coupling approach for this theory to remain valid. As with the other ferropnictides, local AF exchange interactions between the second-nearest neighbours ($J_2$) are expected to be the most relevant for superconductivity in LiFeAs. A natural test is thus to fit the gap amplitude with the global gap function $\cos(k_x)\cos(k_y)$. As shown in Fig. \ref{Fig_111}b, this theory can explain well the anisotropic gap found on the $\beta$ FS. Indeed, the data points are practically perfectly aligned linearly. This simply states that the gap anisotropy on this particular FS comes from the shape of the FS itself. The global gap function also captures the trend of the SC gap size, which shows that larger FSs have smaller SC gap sizes in the ferropnictides. This is particularly true for the $\delta$ FS, which is smaller than the $\gamma$ FS and is thus associated with a larger SC gap amplitude. However, the formula fails to reproduce the SC gap amplitude on the two M-centred electron-like FSs. Interestingly though, the largest discrepancy is observed at the intersection of the two FSs. For this reason, Umezawa \emph{et al.} \cite{UmezawaPRL2012} suggested that this behaviour was related to some hybridization problem. In any case, this observation reveals the limitations of the simplified version of the strong coupling approach presented in the previous sections. 

There is nevertheless an explanation to the gap anisotropy on the $\gamma$ FS that is compatible with the strong coupling scenario. In Yin \emph{et al.} \cite{Yin_ZP_orbital2014}, the calculation of the diagonal part of the SC gap pairing amplitude $\Delta_j(\vec{k})=\langle c^{\dagger}_{k\uparrow,j}c^{\dagger}_{-k\downarrow,j}\rangle$ indicates a strong anisotropy along the M-centred electron-like FSs, with a deep minimum at 45$^{\circ}$ degrees from the $\Gamma$-M direction, exactly as observed experimentally \cite{UmezawaPRL2012,BorisenkoSym2012}. The approach includes the orbital degrees of freedom, an somehow validates the physical intuition of Umezawa \emph{et al.} in attributing the departure from the strong coupling derived global gap function as the result of an hybridization effect \cite{UmezawaPRL2012}, which might involve low-energy physics.

Another notable example of anisotropic gap is found in KFe$_2$As$_2$ ($T_c=3.4$ K) and Ba$_{0.1}$K$_{0.9}$Fe$_2$As$_2$ ($T_c=9$ K), as supported from thermal conductivity data \cite{ReidPRL109,JK_DongPRL2010}. A laser-ARPES study of KFe$_2$As$_2$ reported octet nodes on the $\alpha'$ FS and strong anisotropic SC gap on the $\alpha$ and $\beta$ FSs \cite{Okazaki_Science337}. The average SC gap sizes on these FSs are very small, about 1 and 0.5 meV, respectively.  However, it is important to note that these gap anisotropies may not be representative of the pairing interaction. Indeed, the $k_F$ positions in this study have been determined by using the MDCs, which show a strong overlap of the neighbouring bands. More importantly, the SC gap values were extracted from EDC fits using a Dynes function \cite{Dynes_PRL41}. However, since the EDCs do not show coherent peaks, the fits are largely determined by the position of the leading edge, which can be strongly affected by scattering, as explained in section \ref{section_gap_definition}.

\begin{figure}[!t]
\begin{center}
\includegraphics[width=8 cm]{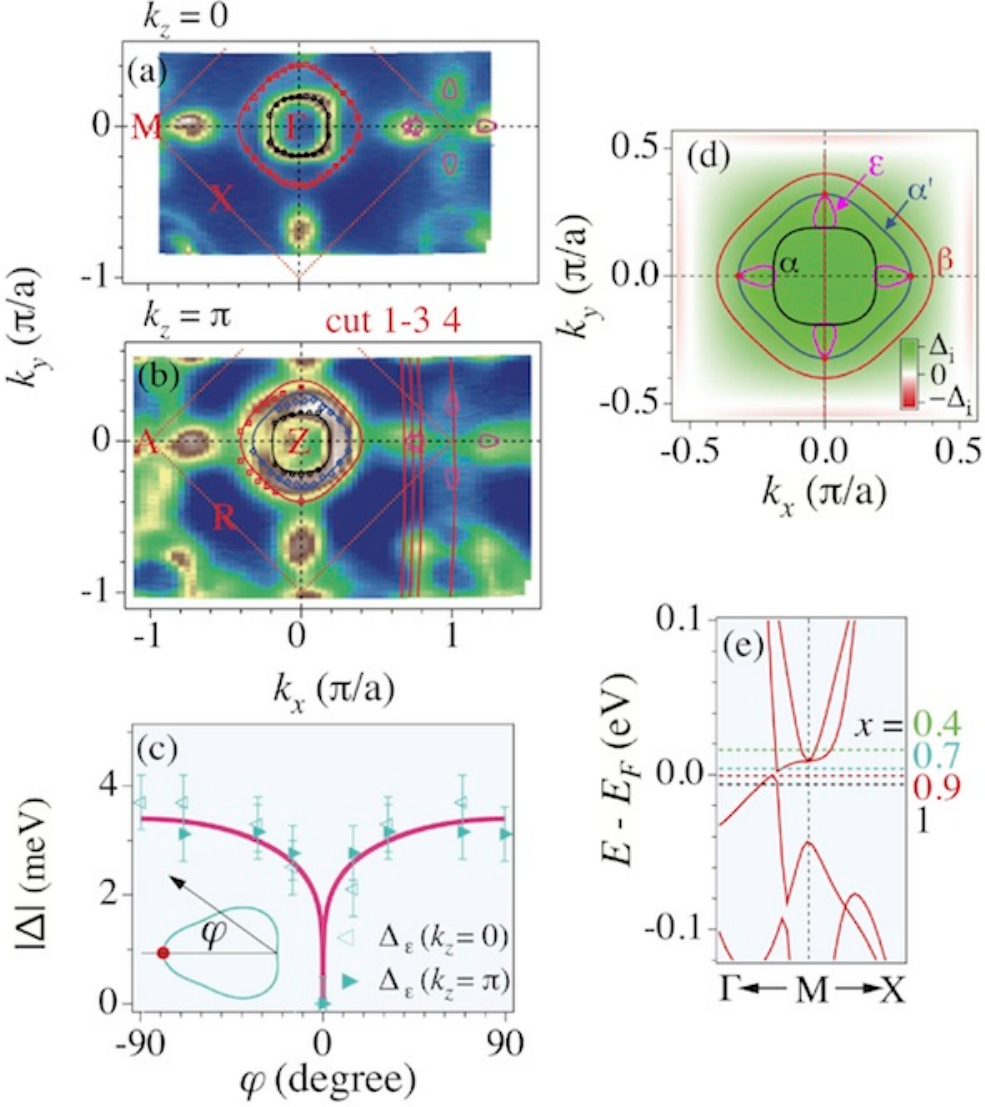}
\end{center}
\caption{\label{Gap_OD}(Colour online). (a) ARPES FS intensity mapping ($\pm 5$ meV integration) recorded with $h\nu= 60$ eV $(k_z = 0)$. (b) Same as (a) but with $h\nu= 60$ eV $(k_z=\pi)$. (c) SC gap size at 0.9 K along the $\varepsilon$ FS as a function of the angle $\varphi$ (defined in the inset). The pink line is a guide for the eye. (d) FS of Ba$_{0.1}$K$_{0.9}$Fe$_2$As$_2$ with the $\varepsilon$ FS pockets shifted by $(-\pi,0)$. The colour scale represents the amplitude of the $\cos(k_x)\cos(k_y)$ global pairing function. (e) LDA band structure calculations from Ref. \cite{G_Xu_EPL2008}, renormalized by a factor 2. The location of the chemical potential is indicated for several doping levels. Reprinted with permission from \cite{Nan_XuPRB88}, copyright \copyright\xspace (2013) by the American Physical Society.}
\end{figure}

In contrast to the laser-ARPES study on KFe$_2$As$_2$, Xu \emph{et al.} reveal rather isotropic SC gaps for the electronic dispersions of the $\Gamma$-centred hole-like FSs of Ba$_{0.1}$K$_{0.9}$Fe$_2$As$_2$ \cite{Nan_XuPRB88}. The SC gap amplitudes in Ba$_{0.1}$K$_{0.9}$Fe$_2$As$_2$ are also significantly larger than in KFe$_2$As$_2$ due to a higher $T_c$. The FS mappings of this material at $k_z=0$ and $k_z=\pi$ are shown in Figs. \ref{Gap_OD}a and \ref{Gap_OD}b, respectively. In addition to the three hole-like FS pockets centred at $\Gamma$, small hole-like M-off-centred $\varepsilon$ FS pockets similar to those observed in KFe$_2$As$_2$ \cite{Sato_PRL2009,Yoshida_JCPS72} are also observed, and no electron-like FS pocket is detected. Interestingly, the SC gap amplitude along the $\varepsilon$ pocket, which could not be measured by laser-ARPES due to its intrinsic momentum field of view limitations, suggests a node at the angle $\varphi=0$ defined in Fig. \ref{Gap_OD}c. As shown in Fig. \ref{Gap_OD}d, the overlap on the $\varepsilon$ FS pocket with the $\cos(k_x)\cos(k_y)$ gap function is inconsistent with the momentum dependence of its SC gap amplitude. Since the tip of the $\varepsilon$ FS is connected by $(0,\pi)$ with the $\alpha'$ FS, Xu \emph{et al.} proposed that low-energy inter-band scattering could be responsible for this peculiar behaviour \cite{Nan_XuPRB88}. 

The recent developments on the determination of the phase of the SC gap, discussed in section \ref{section_phase}, provide alternative scenarios. As shown in Fig. \ref{Gap_OD}e , the $\varepsilon$ pocket emerges as the chemical potential is lowered due to hole doping. This pocket is composed by different orbital characters. While the tip pointing towards $\Gamma$ has a $d_{xy}$ character, the opposite section carries mainly $d_{xz}$ and $d_{yz}$ characters. According to the anti-phase $s_{+-}$ model, the phase of the SC gap on the $d_{xy}$ FSs should be opposite from that formed with $d_{xz}$ and $d_{yz}$ orbitals. Consequently, there must necessarily be a node on the $\varepsilon$ pocket. Since the portion of the pocket having a strong $d_{xy}$ component is small, minimization of energy favours the opening of a large gap on the $d_{xz}/d_{yz}$ section of the $\varepsilon$ pocket and a null gap at the pocket tip with $d_{xy}$ character.	

Another scenario to explain the nodal superconductivity in KFe$_2$As$_2$ emerged with the recent identification by STM and ARPES of a van Hove singularity slightly below $E_F$, located mid-way between the $\Gamma$ and M points \cite{DL_Fang_vHs}. This van Hove singularity is likely to strongly affect the transport properties and it is possibly responsible for the heavy mass behaviour reported for this compound. Interestingly, the zero bias density-of-states measured by STM is not fully gapped in the SC state, even for clean samples. This has been attributed to the fact that the momentum location of the van Hove singularity coincides with the nodal line of the $s_{\pm}$ gap function \cite{DL_Fang_vHs}. It is to note that a van Hove singularity has also been reported for isostructural TlNi$_2$Se$_2$ \cite{Nan_XuTlNi2Se2}, which was claimed to have heavy-electron mass behaviour \cite{H_WangPRL111}, and found to be a nodal superconductor from thermal conductivity measurements \cite{XC_HongPRB90}. ARPES measurements clearly show that both TlNi$_2$Se$_2$ \cite{Nan_XuTlNi2Se2} and KNi$_2$Se$_2$ \cite{Q_Fan_KNi2Se2} are weakly correlated and that the heavy-mass behaviour in these materials is simply due to their particular band structures. 

A circular horizontal node was also reported on the largest hole-like FS at the Z point of BaFe$_2$(As$_{0.7}$P$_{0.3}$)$_2$, which is not expected from the simplified strong coupling model \cite{Y_Zhang_NaturePhys2012}. It was suggested that this node was accidental rather than enforced by symmetry. Although its origin remains a subject of debate, it was tentatively attributed to the strong 3D character of that band at the Z plane, due to the hybridization with the 3$d_{3z^2-r^2}$ orbital. The interpretation of horizontal node was challenged by another ARPES study \cite{Yoshida_Ba122AsP}, in which no horizontal node was found but a strong SC gap anisotropy on the inner M-centred electron-like pocket was proposed. Further ARPES studies are thus necessary to conclude on this particular topic. 

Since ARPES is essentially a surface probe, it is important to make a parallel between the gap structure obtained from ARPES and the one derived from a bulk probe. Thermal conductivity $\kappa(T)$ is arguably the must trustable bulk tool for probing the SC gap structure, or at least to conclude in the presence or absence of nodes. Because the Cooper pairs do not carrier entropy, the observation of a non-zero contribution of the electronic thermal conductivity near the absolute zero temperature (deep into the SC phase) implies that there is at least one point of the FS that is not gapped. Even though thermal conductivity does not probe directly the SC gap structure, the sensitivity to the presence of nodes is very reliable because unless the samples are phase-separated, the observation of nodes does not depend on models or analysis and it is also independent of the presence of impurities.

In agreement with ARPES experiments \cite{Ding_EPL,L_Zhao}, a negligible $\kappa(T\rightarrow 0)/T$ term is measured by thermal conductivity in Ba$_{0.6}$K$_{0.4}$Fe$_2$As$_2$, implying that the FS is not entirely gapped \cite{XG_LuoPRB2009}. Reid \emph{et al.} showed that this is also true for the in-plane and out-of-plane thermal conductivity of the other members of the Ba$_{1-x}$K$_{x}$Fe$_2$As$_2$ series down to $x=0.16$ \cite{Reid2011}, which extends beyond the doping range for which ARPES data of the SC gap have been reported. As discussed above, thermal conductivity reports nodal superconductivity in KFe$_2$As$_2$ \cite{ReidPRL109,JK_DongPRL2010} and TlNi$_2$Se$_2$ \cite{XC_HongPRB90}, in agreement with ARPES. For the Co-doped side of the phase diagram of the 122-ferropnictides, in-plane $\kappa(T\rightarrow 0)/T$ at zero field also suggests the absence of node \cite{L_DingNJP2009,TanatarPRL2010,ReidPRB2010}, which is also consistent with ARPES \cite{Terashima_PNAS2009}. However, accidental nodes have been proposed from $c$-axis measurements data \cite{ReidPRB2010}. Unfortunately, this result cannot be compared directly to the ARPES data of Terashima \emph{et al.} \cite{Terashima_PNAS2009} which are limited to a single photon energy and thus a single $k_z$ value.  

Although the SC gap structure of BaFe$_2$(As$_{1-x}$P$_{x}$)$_2$ is still debated in the ARPES community, nodes have been proposed, which is also consistent with thermal conductivity measurements \cite{YamashitaPRB84}. Tanatar \emph{et al.} report isotropic SC gaps from thermal conductivity in LiFeAs \cite{TanatarPRB84}. Even though some anisotropy is revealed from ARPES \cite{UmezawaPRL2012,BorisenkoSym2012}, the SC gap structure obtained from ARPES is far from a nodal structure. There is one noticeable case where ARPES has not been able to identify a node suggested by thermal conductivity, namely isovalent-substituted Ba(Fe$_{1-x}$Ru$_x$)$_2$As$_2$. While ARPES data at various photon energies are consistent with an isotropic SC gap \cite{Nan_XuPRB87}, nodal superconductivity is suggested from thermal conductivity measurements \cite{X_Qiu_PRX2012}. The reason for this discrepancy is still unknown, but possibilities include phase-separation since this material usually exhibits a small SC volume fraction, which would affect the thermal conductivity measurements, or accidental nodes on parts of the FS that have not been probed directly by ARPES.

In short, within error bars and besides potential technical issues, the agreement between ARPES and thermal conductivity, a highly trusted bulk probe of nodes in the SC gap, is rather encouraging and reinforces the conclusions derived from systematic recording by ARPES of the SC gap structure directly in the momentum space.

\section{Orbital effects}
\label{section_orbital}

The previous section on the SC gap anisotropy in some materials clearly suggests the relevance of the orbital degree of freedom. In fact, this is not a surprise in the context of the strong coupling approach. Indeed, the local orbital configuration is largely responsible for the local moment, and thus the orbital and spin degrees of freedoms are necessarily strongly coupled \cite{CC_LeePRL103}. The orbital configuration is also intimately related to the exchange and hopping parameters at the centre of the strong coupling description \cite{WG_YinPRB79}. Such necessary coupling between the spin and orbital degrees of freedom has also been pointed out using weak coupling approaches \cite{Fernandes_nphys10}, and there is a growing consensus on the importance of the orbital degree of freedom, at least for the description of some physical behaviours. The orbital fluctuations have been proposed to be the cause of the structural phase transition occurring at high temperature in many parent compounds of Fe-based superconductors \cite{CC_LeePRL103,W_LvPRB82}. It has also been suggested that the orbital fluctuations are closely related to the electronic nematicity and giant magnetic anisotropy found experimentally in some Fe-based compounds \cite{Chuang_Science327,Chu_Science329,M_YiPNAS2011,Y_Zhang_PRB85,M_Yi_NJP14,Hung_PRB85,Kasahara_Nature486, Shimojima_PRB89}. Finally and not the least, orbital fluctuations have been proposed for the pairing mechanism itself \cite{Kontani_PRL104,S_Zhou_PRB84}. 

Despite predictions, direct connections between orbital fluctuations and Fe-based superconductivity are not easy to find experimentally. A recent ARPES study focussing mainly on LiFeAs showed a direct relationship between ferro-orbital fluctuations and superconductivity \cite{H_Miao_PRB89}. In addition of having a structure that leaves non-polar cleaved surfaces, LiFeAs is free of structural and magnetic transitions \cite{XC_WangSSC2008}, and is thus perfectly suited to investigate the possible correlation between orbital fluctuations and superconductivity. Due to the four-fold symmetry of the system, one would assume the $\alpha$ and $\alpha'$ bands, which origin from the $d_{xz}$ and $d_{yz}$ orbitals, to be degenerate at the $\Gamma$ point. Interestingly, that is not what is observed experimentally. The removal of the degeneracy at the $\Gamma$ point implies directly, whatever the cause of this phenomenon, a misbalance in the occupation of the $d_{xz}$ and $d_{yz}$ orbitals. In the absence of long-range ordering, one must conclude that the system shows ferro-orbital fluctuations \cite{H_Miao_PRB89}. 

\begin{figure}[!t]
\begin{center}
\includegraphics[width=14 cm]{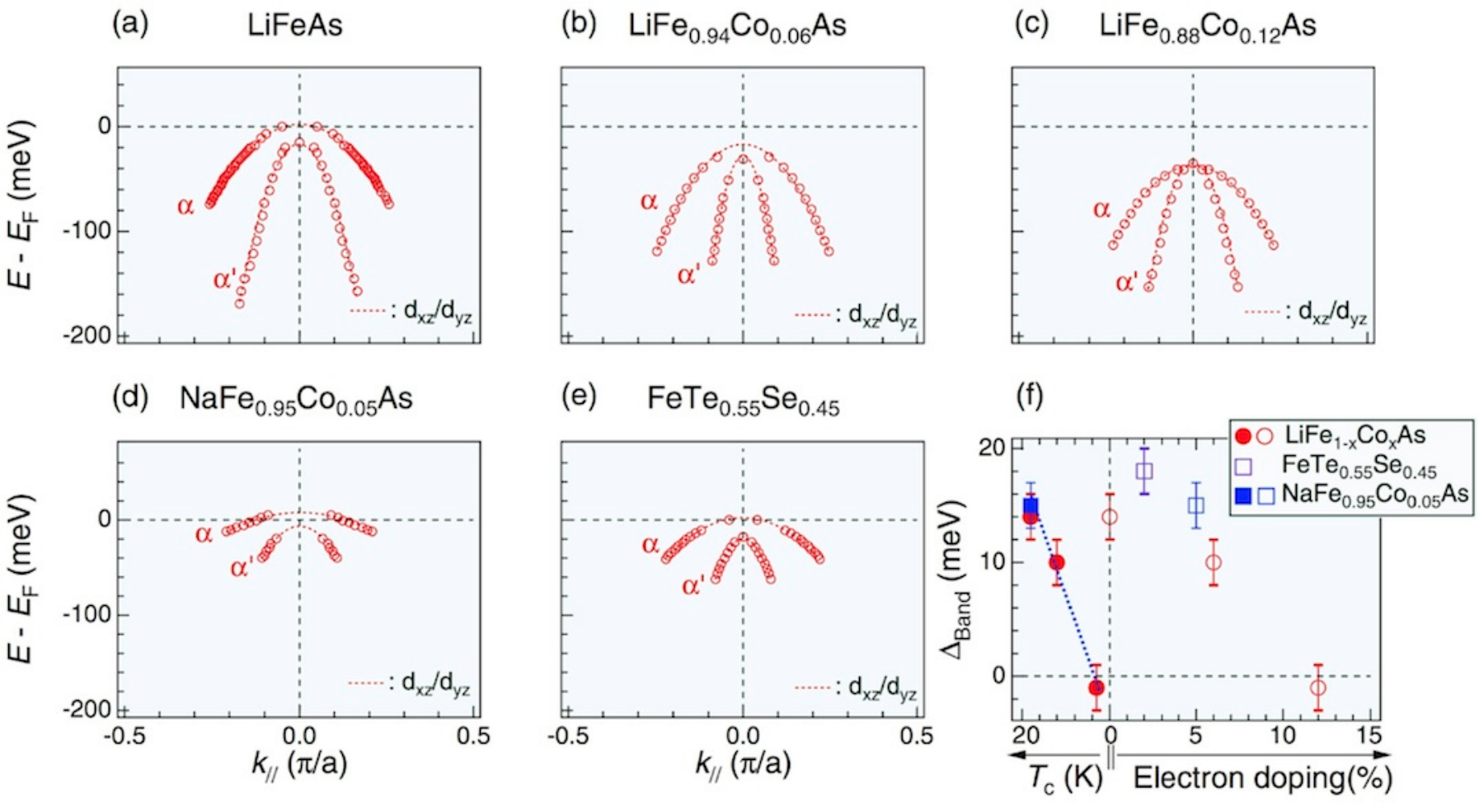}
\end{center}
\caption{\label{LiFeAs_ferro}(Colour online). (a) - (c), Extracted band dispersion of the $d_{xz}/d_{yz}$ bands in LiFeAs, LiFe$_{0.94}$Co$_{0.06}$As and LiFe$_{0.88}$Co$_{0.12}$As \cite{H_Miao_PRB89}, respectively. (d) and (e), Extracted band dispersion of NaFe$_{0.95}$Co$_{0.05}$As \cite{ZH_LiuPRB84} and FeTe$_{0.55}$Se$_{0.45}$ \cite{H_Miao_PRB2012}, respectively. Red dashed curves are parabolic fits. (f), Doping and $T_{c}$ dependence of $\Delta_{\textrm{band}}$. The open and plain symbols refer to the doping (bottom right) and $T_{c}$ (bottom left) axes. Reprinted with permission from \cite{H_Miao_PRB89}, copyright \copyright\xspace (2014) by the American Physical Society.}
\end{figure}

We show in Fig. \ref{LiFeAs_ferro} the extraction of the electronic dispersion of the $\alpha$ and $\alpha'$ bands in LiFeAs and other Fe-based superconductors for which the top of these two bands locate near above or below $E_F$. In particular, the top row shows the situation in LiFe$_{1-x}$Co$_x$As for $x=0$, $x=0.06$ and $x=0.12$, which have $T_c$'s of 18, 10 and 4 K, respectively. The top of the $\alpha$ band, which is located slightly above $E_F$ in LiFeAs, sinks below $E_F$ following the introduction of carriers by the partial substitution of Fe by Co. The experimental data show clearly that the top of the $\alpha$' band is located 14 meV below that of the $\alpha$ band in LiFeAs. Interestingly, this splitting decreases as the Co content is increased to $x=0.06$, and within error bars the tops of the $\alpha$ and $\alpha'$ bands are degenerate at $x=0.12$.

The observation of the removal of the $d_{xz}/d_{yz}$ degeneracy at $\Gamma$ is also observed in other materials. As shown in Fig. \ref{LiFeAs_ferro}d, a splitting $\Delta_{band}$ of 15 meV is also recorded in NaFe$_{0.95}$Co$_{0.05}$As, which has the same $T_c$ as LiFeAs. The left side of Fig. \ref{LiFeAs_ferro}f suggests that there is a direct correlation in the 111-ferropnictide family between $T_c$ and the size of the $\alpha/\alpha'$ splitting \cite{H_Miao_PRB89}. An even larger splitting of 18 meV is observed in FeTe$_{0.55}$Se$_{0.45}$, as illustrated in Fig. \ref{LiFeAs_ferro}e, although this latter splitting seems not to follow the same scaling as in the 111-ferropnictides.

Miao \emph{et al.} \cite{H_Miao_PRB89} showed that while the dispersion of the $\alpha'$ band is unaffected with temperature increasing, the top of the $\alpha$ band shifts towards the higher binding energies, thus reducing the splitting between the $\alpha'$ and $\alpha$ bands, which is almost closed at 250 K in LiFeAs. More importantly, the splitting persists below $T_c$, indicating the coexistence of ferro-orbital fluctuations and superconductivity. The correlation between $T_c$ and $\Delta_{band}$ persisting in the SC state indicates that the fluctuations of the ferro-orbital order is intimately connected to superconductivity. This contrasts with the observation by NMR of enhanced low-energy AF correlations as $T_c$ decreases in LiFe$_{1-x}$Co$_x$As \cite{YM_Dai_soon}.

Since the $d_{xz}$ and $d_{yz}$ orbitals have strongly anisotropic quasi-one-dimensional (1D) hopping integrals, we can refine our understanding of the ferro-orbital fluctuations using a a simple quasi-1D model with local ferro-orbital fluctuations represented by an Ising field \cite{CH_LinPRL107}. Interestingly, indications of the fluctuating ferro-orbital fluctuations other than the broadening of the quasi-particles appear clearly only when the spatial correlations decay with a power-law. It is also important to note that although spin orbit coupling can in principle remove the degeneracy at $\Gamma$ while preserving tetragonal symmetry \cite{Cvetkovic_PRB88}, the strong doping dependence and the temperature evolution of $\Delta_{band}$ are inconsistent with this scenario. 

Although this review is mainly devoted to the measurements of SC gaps by ARPES, this chapter would not be complete without commenting briefly on the importance of the electronic correlations for the electron pairing in Fe-based superconductors, particularly in the context of the strong coupling approach. Indeed, the electronic correlations in the Fe-based superconductors are not negligible and lead to the renormalization of the electronic band structure by typical factors of 2-5 over an energy range of 1 eV or more \cite{RichardRoPP2011}, which cannot be explained uniquely by low-energy physics. As with the cuprates, the Fe-based superconductors share an electronic structure in which the bands near $E_F$ mainly derive from $3d$ orbitals. However, the electronic transport is not directly due to the overlap between these $d$ orbitals, but rather by super-exchange processes through intermediate atoms that control the hopping parameters (O in the case of the cuprates and pnictide or chalcogenide atoms in the case of the Fe-based superconductors). In other words, even the electrons said ``itinerant" are in fact partly localized on the sites of the $d$ orbitals, which can be viewed as the origin of band renormalization. This partial localization has an even more important consequence. Indeed, the electrons are consequently very sensitive to the local atomic configurations, which can include several parameters such as the local spin configuration, the local orbital configuration and of course the local Coulomb interactions. The hopping between two neighbour sites thus depends critically on their respective local configurations, which is essentially what short-range electronic correlations mean. Conceptually, the fluctuations of these local parameters can play a role that is the analogue of the charge fluctuations in conventional superconductors, which are driven by the vibrations of the atomic lattice. 

In order to optimize the pairing strength, it is necessary to preserve the electronic correlations while preventing them from being too strong, which would mean that electrons would become too localized and thus unable to contribute efficiently to electronic transport. The isovalent-substituted Ba(Fe$_{1-x}$Ru$_{x}$)$_2$As$_2$ system illustrates well the connexion between high $T_c$ values and electronic correlations. The substitution of Fe $3d$ orbitals by Ru $4d$ orbitals first leads to the suppression of the long-range AFM order and to the emergence of superconductivity \cite{Sharma_prb2010,Albenque_prb2010,Thaler_PRB2010Ru}. While this effect was first attributed to the reduction of the correlation effects due to the introduction of extended $4d$ orbitals by an ARPES investigation \cite{Brouet_PRL105}, the observation of relatively constant Fermi velocities as a function of doping led another group to conclude that the principal effect of this substitution is to dilute the magnetic structure \cite{Dhaka_prl2011}. A third ARPES study over a wider range of substitution demonstrated that while the Fermi velocities are approximately constant up to a Ru content of $x=0.3-0.4$, which coincides with the optimal $T_c$, a sudden increase of the Fermi velocities and an enhanced 3D character take place at higher substitution levels, as $T_c$ starts to decrease, thus suggesting the importance of the electronic correlations for maximizing $T_c$ \cite{Nan_XuPRB86}. 

There is also increasing theoretical evidence for the role played by the Hund's rule coupling and the filling of the $3d$ band in tuning the strength of the electronic correlations \cite{HauleNJP11,AichhornPRB82,LiebschPRB82,IshidaPRB81}. In an ARPES study of BaCo$_2$As$_2$, Xu \emph{et al.} \cite{Nan_XuPRX3} showed that at the first order BaCo$_2$As$_2$ could be used to visualize states corresponding to unoccupied states in the 122-ferropnictides. However, as later confirmed by another ARPES study \cite{Dhaka_prb87}, the electronic structure of this material is only slightly renormalized as compared to its 122-ferropnictide cousins \cite{Nan_XuPRX3}. Interestingly, this study indicated that the $\beta$ band, which origins mainly from the $d_{xy}$ orbital, was twice as much renormalized as the others in Ba$_{0.6}$K$_{0.4}$Fe$_2$As$_2$ than the others, a clear sign of the orbital dependence of the electronic correlations. Based on a good agreement between the experimental data and LDA+DMFT (dynamic mean-field theory) calculations of the electronic band structure of BaCo$_2$As$_2$, the effect of band filling was investigated theoretically by comparing BaFe$_2$As$_2$ with an effective system consisting in BaFe$_2$As$_2$ with an additional electron per Fe. These calculations on this artificial system give results almost identical to the ones obtained in BaCo$_2$As$_2$, indicating a linear behaviour of the self-energy as a function of the Matsubara frequency that contrasts to the nearly square-root behaviour observed in BaFe$_2$As$_2$, strongly suggesting that the reduction of the electronic correlations in BaCo$_2$As$_2$ is essentially due to electron filling in the presence of a large Hund's rule coupling term \cite{Nan_XuPRX3}.

\section{Concluding remarks}

The physical phenomena surrounding us are described by mathematical laws that we can access through an iterative process of experiments and mathematical modelling. The best model is usually the one that captures most of the physics without unnecessary complications, and there is no apparent reason why this rule of thumb should not apply to superconductivity. The understanding of conventional superconductivity that emerged from the basic concept of Cooper pair and the subsequent BCS and Eliashberg theories has long been regarded as one the greatest achievements in condensed matter physics. This concept can easily be explained with plain words: the interaction between one electron and a lattice may induce a ``dynamical deformation" of that lattice favouring the attraction of another electron, which can be seen as a retarded effective electron-electron attraction. In conventional superconductors, it is the ionic charge lattice that is deformed over a relatively long distance compared with a unit cell. After one worked out the electronic and phonon structures properly this picture is valid for \emph{all} conventional superconductors.

With the discovery of the cuprate superconductors, the universality of phonon-mediated superconductivity was seriously challenged. Later, after the discovery of Fe-based superconductivity, most of the community agreed that ionic charge lattice cannot provide the proper glue for high-temperature superconductivity, and most hints now point towards the importance of antiferromagnetism. Yet, disagreement persist as to how antiferromagnetism leads to electron pairing. At the core of this debate, two philosophies are facing each other: on one hand, the weak coupling scenarios state that the Fermi surface controls directly the pairing of electrons; on the other hand, the strong coupling theories promote short-range interactions as the key players for the electron pairing.

In this topical review, we demonstrated using ARPES that the complicated evolution of the FS of the Fe-based superconductors with doping and crystal structure is incompatible with \emph{any} weak coupling theory for describing the electron pairing. Indeed, the price to pay for maintaining these scenarios alive is to conclude that there are several mechanisms for Fe-based superconductivity, even for a single crystal structure. For example, although they share the same basic crystal structure, optimally-doped Ba$_{0.6}$K$_{0.4}$Fe$_2$As$_2$, KFe$_2$As$_2$ (hole pockets only) and the 122-ferrochalcogenides (electron pockets only) have significantly different FSs, and the latter two do not even have possibility for electron-hole quasi-nesting. In the context of inter-pocket and intra-pocket FS interactions, one must consider at least 3 different pairing mechanisms for the 122 system, a serious step away from simplification and universality. 

In contrast, we showed in this topical review how consistent, robust and yet so simple is the strong coupling picture in describing the ARPES results related to the pairing of electrons in the Fe-based superconductors: 

\begin{enumerate}
\item The $J_1$-$J_2$-$J_3$ model can be used to characterize the spin-wave dispersion from inelastic neutron scattering experiments of the magnetic parent compounds, and thus to parameterize the local exchange interactions.
\item The relevant antiferromagnetic local exchange parameters lead to simple form factors when expressed in the momentum space that can be \emph{mapped out over the entire first BZ} of the SC materials considered.  
\item The pairing amplitude at a particular momentum $k_F$ depends only on its \emph{absolute position} in the momentum space. 
\end{enumerate}

In agreement with neutron experiments and regardless of the details of the FS, we showed evidence from ARPES gap measurements for a leading $s$-wave pairing term in $J_2$ \emph{for all} the Fe-based systems that we studied. Although modulations from the simple $\cos k_x\cos k_y$ global gap function are observed when inter-layer interactions are important or when the $J_3$ parameter is non-negligible, we can claim from the strong coupling approach that \emph{the same pairing mechanism applies to all the Fe-based superconductors}, which is a significant step towards simplification and universality in Fe-based superconductivity. In fact, this is probably an even bigger step towards the universality of the pairing mechanism for a much broader class of unconventional superconductors that includes the cuprates and the heavy-fermion materials, as the same recipe enumerated above can be applied to these systems as well. Of course, one should caution that this does not mean that all physical properties derive from the strong coupling approach. On the contrary, FS effects remain important players in the physics of the Fe-based materials, even for superconductivity-related issues such as pair breaking.

What does our main conclusion on the validity of the strong coupling approach means physically? Somehow, it means that unconventional superconductivity itself is not that much different from conventional superconductivity. Instead of an effective interaction between electrons mediated by the ionic charge lattice, the effective interaction is now provided by the lattice of the local moments that modulates the exchange interactions, or by any local property that is directly correlated with local moments. In strongly correlated electron systems, the electron is very sensitive to local parameters such as the local moment. In analogy with the ionic charge lattice for conventional superconductivity, it is the ``dynamical deformation" of the local moment lattice (spin fluctuations) that assures the paring in unconventional superconductors. Even though this simple picture may need refinement for quantitative predictions, it certainly contains the key elements for a final understanding of superconductivity in Fe-based materials and other unconventional systems.

\ack
We acknowledge Jiangping Hu for useful discussions. This work was supported by grants from MOST (2010CB923000,  2011CBA001000, 2011CBA00102, 2012CB821403 and 2013CB921703) and NSFC (11004232, 11034011/A0402, 11234014 and 11274362) from China.

\section*{References}
\bibliographystyle{unsrt} 
\bibliography{biblio_long}

\end{document}